\documentclass{lmcs} %
\pdfoutput=1

\usepackage{lastpage}
\lmcsdoi{17}{4}{9}
\lmcsheading{}{\pageref{LastPage}}{}{}%
{Jul.~15,~2020}{Nov.~16,~2021}{}

\usepackage[utf8]{inputenc}


\usepackage{hyperref}

\usepackage{mathtools}

\usepackage[pagewise]{lineno}
\usepackage{graphicx}
\usepackage{xspace}
\usepackage{amssymb}
\usepackage{soul}
\usepackage{tikz}
\usetikzlibrary{arrows,decorations.pathmorphing,positioning,decorations.pathreplacing, shapes.geometric, patterns}
\usepackage{subcaption}

\newtheorem{innercustomthm}{Theorem}
\newenvironment{customthm}[1]
  {\renewcommand\theinnercustomthm{#1}\innercustomthm}
  {\endinnercustomthm}
\newtheorem{innercustomcor}{Corollary}
\newenvironment{customcor}[1]
  {\renewcommand\theinnercustomcor{#1}\innercustomcor}
  {\endinnercustomcor}

\newcommand{\Parity}{\myproblem{Parity}}

\newcommand{\RelName}[1]{\mtext{#1}}
\newcommand{\tpl}{\bar}

\newcommand{\mtext}[1]{\textsc{#1}}
\newcommand{\ins}{\mtext{ins}\xspace}
\newcommand{\del}{\mtext{del}\xspace}

\newcommand{\schema}{\ensuremath{\sigma}\xspace}

\newcommand{\query}{\ensuremath{q}}
\newcommand{\db}{\ensuremath{\calI}\xspace}%
\newcommand{\inp}{\ensuremath{\calI}\xspace}
\newcommand{\aux}{\ensuremath{\calA}\xspace}%

\newcommand{\ans}{\RelName{Ans}}

\newcommand{\BIT}{\ensuremath{\text{\upshape BIT}}}

\newcommand{\kpower}[2][k]{\ensuremath{\binom{#2}{#1}}}%

\newcommand{\N}{\ensuremath{\mathbb{N}}}

\newcommand{\bigO}{\ensuremath{\mathcal{O}}}

\newcommand{\df}{\ensuremath{\mathrel{\smash{\stackrel{\scriptscriptstyle{
    \text{def}}}{=}}}} \;}

\newcommand{\uf}[3]{\ensuremath{\varphi_{#1}^{#2}(#3)}}    

\DeclareMathOperator{\indeg}{in-deg}

\newcommand  {\myclass} [1]  {\ensuremath{\textsf{\upshape #1}}}

\newcommand{\StaClass}[1]{\myclass{#1}\xspace}

\newcommand{\DynClass}[1]{\myclass{Dyn#1}\xspace}

\newcommand  {\myproblem} [1] {\normalfont{\textsc{#1}}\xspace}

\newcommand     {\NL}   {\StaClass{NL}}

\newcommand     {\AC}   {\StaClass{AC}}
\newcommand     {\ACC}   {\myclass{ACC}}
\newcommand{\ACz}{\mbox{\myclass{AC}$^0$}\xspace}

\newcommand{\qDynAC}{\ensuremath{\text{q-}\DynClass{AC}}}
\newcommand     {\Sym}   {\myclass{Sym}^+}

\newcommand{\FO}{\StaClass{FO}}
\newcommand{\FOparity}{\StaClass{FO+Parity}}
\newcommand{\MSO}{\StaClass{MSO}}

\newcommand{\CQ}[1][]{\StaClass{CQ}}
\newcommand{\UCQ}[1][]{\StaClass{UCQ}}
\newcommand{\CQneg}[1][]{\StaClass{CQ\ensuremath{^{\mneg}}}}
\newcommand{\UCQneg}[1][]{\StaClass{UCQ\ensuremath{^{\mneg}}}}

\newcommand{\mneg}{\neg} %

\newcommand{\DynProp}{\DynClass{Prop}}

\newcommand{\DynFO}{\DynClass{FO}}
\newcommand{\DynAC}{\DynClass{AC}}

\newcommand{\qAC}   {\ensuremath{\text{q-}\myclass{AC}}}
\newcommand{\qACC}   {\ensuremath{\text{q-}\myclass{ACC}}}

\newtheorem*{question*}{Question}
\newtheorem*{openquestion*}{Open question}

\newenvironment{proofsketch}{\begin{proof}[Proof sketch.]}{\end{proof}}

\newenvironment{proofof}[1]{\begin{proof}[Proof (of #1).]}{\end{proof}}

\providecommand {\calA}      {{\mathcal A}\xspace}
\providecommand {\calB}      {{\mathcal B}\xspace}
\providecommand {\calC}      {{\mathcal C}\xspace}

\providecommand {\calI}      {{\mathcal I}\xspace}

\providecommand {\calN}      {{\mathcal N}\xspace}

\providecommand {\calP}      {{\mathcal P}\xspace}

  \newcommand{\changeRule}[3]{\textbf{on change}\ #1\ \textbf{update}\ #2\ \textbf{as}\ #3}

\newcommand{\prog}{\ensuremath{\calP}\xspace}

\newcommand{\inpSchema}{\ensuremath{\schema_{\text{in}}}\xspace}
\newcommand{\auxSchema}{\ensuremath{\schema_{\text{aux}}}\xspace}

\newcommand{\Nin}{\ensuremath{\calN^\leftarrow}} 
\newcommand{\NoutE}{\ensuremath{\calN_\exists^\rightarrow}} 
\newcommand{\NoutA}{\ensuremath{\calN_\forall^\rightarrow}} 
\newcommand{\Ncunc}{\ensuremath{\calN^{\bullet \circ}}} 
\newcommand{\Ncorunc}{\Ncunc} 
\newcommand{\xor}{\oplus}
\newcommand{\parityexists}{\myproblem{ParityExists}}

\newcommand{\parityexistsdeg}[1]{\ensuremath{\parityexists_{\deg\leq#1}}}

\tikzstyle{mnode}=[
  circle,
  fill=\mnodefillcolor, 
  draw=\mnodedrawcolor,
  minimum size=6pt, 
  inner sep=0pt
]

\tikzstyle{dDashedEdge}=[
  -latex', %
  thick, 
  shorten >=3pt, 
  shorten <=3pt,
  draw=black!80,
  dashed
]

\tikzstyle{dEdge}=[
  -latex', %
  thick, 
  shorten >=1pt, 
  shorten <=1pt,
  draw=black!80,
]

\tikzstyle{uEdge}=[
  thick, 
  shorten >=3pt, 
  shorten <=3pt,
  draw=black!80,
]
\tikzstyle{mline}=[
  draw=black,
  inner sep=0.2cm,
  rounded corners=5pt,
  thick
]

\newcommand{\mnodedrawcolor}{black!80}
\newcommand{\mnodefillcolor}{black!20}

\pgfdeclarelayer{background}
\pgfsetlayers{background,main}%

\newcommand{\statementDynFOlogn}{$\parityexistsdeg{\log n}$ can be maintained in \DynFO with binary auxiliary relations in the presence of a linear order and \BIT.}  
\newcommand{\statementDynFOdegk}{$\parityexistsdeg{k}$ can be maintained in \DynFO with unary auxiliary relations in the presence of a linear order, for every $k \in \N$.}

\begin{document}

\title{Dynamic Complexity of Parity Exists Queries}

\author[N.~Vortmeier]{Nils Vortmeier\rsuper{a}}	%
\address{University of Zurich, Switzerland}	%
\email{nils.vortmeier@uzh.ch}  %

\author[T.~Zeume]{Thomas Zeume\rsuper{b}}	%
\address{Ruhr University Bochum, Germany}	%
\email{thomas.zeume@rub.de}  %

\begin{abstract}
Given a graph whose nodes may be coloured red, the parity of the number of red nodes can easily be maintained with first-order update rules in the dynamic complexity framework \DynFO of Patnaik and Immerman.
Can this be generalised to other or even all queries that are definable in first-order logic extended by parity quantifiers?
We consider the query that asks whether the number of nodes that have an edge to a red node is odd. Already this simple query of quantifier structure parity-exists is a major roadblock for dynamically capturing extensions of first-order logic.

We show that this query cannot be maintained with quantifier-free first-order update rules, and that variants induce a hierarchy for such update rules with respect to the arity of the maintained auxiliary relations. Towards maintaining the query with full first-order update rules, it is shown that degree-restricted variants can be maintained.
\end{abstract}

\maketitle

\section{Introduction}
\label{section:introduction}
The query \Parity{} --- given a unary relation $U$, does $U$ contain an odd number of elements? --- cannot be \emph{expressed} in first-order logic, even with arbitrary numerical built-in relations~\mbox{\cite{Ajtai83, FurstSS84}}. 
However, it can easily be \emph{maintained} in a dynamic scenario where single elements can be inserted into and removed from $U$, and helpful information for answering the query is stored and updated by first-order definable update rules upon changes. Whenever a new element is inserted into or an existing element is removed from $U$, then a stored bit $P$ is flipped\footnote{This bit is preserved if a change re-inserts an element that already is in $U$, or tries to delete an element that is not in $U$.}. 
In the dynamic complexity framework by Patnaik and Immerman \cite{PatnaikI97} this can be expressed by the following first-order update rules:
         \begin{align*}
&\textbf{on insert } a \text{ into } U  \textbf{ update }  P \textbf{ as }  (\neg U(a) \wedge \neg P) \; \vee \; (U(a) \wedge P)\\
&\textbf{on delete } a \text{ from } U  \textbf{ update }  P \textbf{ as }  (U(a) \wedge \neg P) \; \vee \;  (\neg U(a) \wedge P)
         \end{align*} 
This simple program proves that \Parity is in the dynamic complexity class $\DynFO$ which contains all queries that can be maintained via first-order formulas that use (and update) some additional stored auxiliary relations.

Motivated by applications in database theory and complexity theory, the class $\DynFO$ has been studied extensively in the last three decades. In database theory, it is well-known that first-order logic corresponds to the relational core of SQL (see, e.g., \cite{AbiteboulHV95}). Thus, if a query can be maintained with first-order update rules then, in particular,  it can  be updated using SQL queries. From a complexity theoretic point of view, first-order logic with built-in arithmetic corresponds to the circuit complexity class uniform $\AC^0$ \cite{BarringtonIS90}. Hence queries in $\DynFO$ can be evaluated in a highly parallel fashion in dynamic scenarios.

The focus of research on $\DynFO$ has been its expressive power. The parity query is a first witness that \DynFO is more expressive than \FO\ (the class of queries expressible by first-order formulas in the standard, non-dynamic setting), but it is not the only one. Further examples include the reachability query for general directed graphs \cite{DattaKMSZ18}, another textbook query that is not in \FO but complete for the complexity class \NL, which can be characterised (on ordered structures) by the extension of first-order logic with a transitive closure operator.
On (classes of) graphs with bounded treewidth, \DynFO includes all queries that can be defined in monadic second-order logic \MSO \cite{DattaMSVZ19}, which extends first-order logic by quantification over sets. In particular, \DynFO contains all \MSO-definable Boolean queries on strings, that is, all regular languages. For strings, the first-order update rules do not even need any quantifiers \cite{GeladeMS12}, proving that regular languages are even in the dynamic complexity class \DynProp which is defined via quantifier-free first-order update~rules.

These examples show that in the dynamic setting, first-order logic can, in some cases, sidestep quantifiers and operators which it cannot express statically: parity and set quantifiers, as well as transitive closure operators.
Immediately the question arises whether first-order update rules can dynamically maintain \emph{all} queries that are statically expressible in extensions of first-order logic by one of these quantifiers or operators.
Note that this does not follow easily, for instance, from the result that the \NL-complete reachability query is in \DynFO, because the notions of reductions that are available in the dynamic setting are too weak~\cite{PatnaikI97}.

The extension \FOparity of first-order logic by parity quantifiers is the natural starting point for a more thorough investigation of how $\DynFO$ relates to extensions of $\FO$, as it is arguably the simplest natural extension that extends the expressive power.
Unfortunately, however, a result of the form $\FOparity \subseteq \DynFO$ is not in sight\footnote{Formally one has to be a little more precise. If one does not allow an appropriate initialisation mechanism, one cannot express the query ``The size of the domain is even.'' in $\DynFO$, which implies $\FOparity \not\subseteq \DynFO$ for this variant of \DynFO.
However, if the initialisation is powerful enough to answer the query for the initial input, maintenance under changes is trivial, as they are not allowed to change the domain. 

We aim to study the ability of dynamic programs to maintain queries under changes, we are less interested in inexpressibility results that crucially rely on weaknesses of the initialisation. Therefore we are interested in results of the form $\FOparity \subseteq \DynFO$ for domain-independent queries, that is, queries whose result does not change when isolated elements are added to the domain. As we only consider initial input structures with empty relations, and so initial inputs only differ in their domain, the initial query answer for domain-independent queries is the same for all possible initial input structures, and therefore trivial to obtain regardless of the power of the initialisation.}. While \Parity\ is in $\DynFO$, already for slightly more complex queries expressible in $\FOparity$ it seems not to be easy to show that they are in \DynFO.
In this paper we are particularly interested in the following generalisation of the parity query:

\begin{quote}
 \parityexists: Given a graph whose nodes may be coloured red. Is the number of nodes connected to a red node odd? Edges can be inserted and deleted; nodes can be coloured or uncoloured. 
\end{quote}

As it is still unknown whether \parityexists is in \DynFO, this query is a roadblock for showing that \DynFO captures (large subclasses of) \FOparity. For this reason we study the dynamic complexity of \parityexists. We focus on the following two directions: (1) its relation to the well-understood quantifier-free fragment $\DynProp$ of $\DynFO$, and (2) the dynamic complexity of degree-restricted variants.

The update rules given above witness that \Parity is in \DynProp. %
We show that this is not the case any more for \parityexists.
  \begin{thm}\label{theorem:parityexists:dynprop}
      $\parityexists \not\in \DynProp$. 
  \end{thm}

A fine-grained analysis of the quantifier-free complexity is the main contribution of this paper, which also implies Theorem~\ref{theorem:parityexists:dynprop}. Let $\parityexistsdeg{k}$ be the variant of the $\parityexists$ query that asks whether the number of nodes that have both an edge to a red node and degree at most $k$ is odd, for some fixed number $k \in \N$.

\begin{thm}\label{theorem:parityexists:dynpropdegk}
      $\parityexistsdeg{k}$ can be maintained in \DynProp with auxiliary relations of arity $k$, for any $k \geq 3$, but not with auxiliary relations of arity $k-1$, even on graphs with in-degree at most $k$.
  \end{thm}
  
This result actually has an impact beyond the lower bound given by Theorem~\ref{theorem:parityexists:dynprop}. It clarifies the structure of \DynProp, as it shows that auxiliary relations with higher arities increase the expressive power of quantifier-free update formulas even on graph structures.

Already Dong and Su showed that \DynFO has an arity hierarchy \cite{DongS98}, i.e., that for each $k \in \N$ there is a query $\query_k$ that can be maintained using first-order update rules and $k$-ary auxiliary relations, but not using $(k-1)$-ary auxiliary relations. The query $\query_k$ given by Dong and Su \cite{DongS98} is a $k$-ary query that is evaluated over a $(6k+1)$-ary relation $T$ and returns all $k$-ary tuples $\tpl a$ such that the number of $(5k+1)$-ary tuples~$\tpl b$ with $(\tpl a, \tpl b) \in T$ is divisible by~$4$. Dong and Su ask whether the arity of the relation $T$ can be reduced to $3k$, $k$, or even to~$2$. Their question for reducing it below $3k$ was motivated by a known reduction of the arity to $3k+1$ \cite{DongZ00}.

An arity hierarchy for \DynProp, though again only for input relations whose arity depends on $k$, follows from the observation that the query $q_k$ can be maintained with quantifier-free update rules. Some progress towards an arity hierarchy for queries over a fixed schema was made by Zeume and Schwentick \cite{ZeumeS15}, who separated the arities up to $k=3$ for Boolean queries on graphs. If only insertions are allowed, then \DynProp is known to have an arity hierarchy for Boolean graph queries \cite{Zeume17}. 

An arity hierarchy for quantifier-free update rules and Boolean graph properties is now an immediate consequence of Theorem~\ref{theorem:parityexists:dynpropdegk}, in connection with the separation results for $k \leq 3$ \cite{ZeumeS15}.
 \begin{cor}\label{corollary:arityhierarchy}
    \DynProp has a strict arity hierarchy for Boolean graph queries. 
 \end{cor}
We note that such an arity hierarchy does \emph{not} exist for \DynProp when we consider not graphs as inputs but strings. Gelade et al.\ show that the class of Boolean queries on strings that can be maintained in \DynProp are exactly the regular languages, and that every such language can be maintained with binary auxiliary relations \cite{GeladeMS12}. So, relations of higher arity are never necessary in this case.

With respect to $\DynFO$, we cannot answer the question whether $\parityexists \in \DynFO$, but we can generalise the upper bound of Theorem \ref{theorem:parityexists:dynpropdegk} to restrictions beyond fixed numbers $k$, at least if the update formulas have access to additional built-in relations. 
Let $\parityexistsdeg{\log n}$ be the query that asks for the parity of the number of nodes that are connected to a red node and have degree at most $\log n$, where $n$ is the number of nodes of the graph.
The binary $\BIT$ predicate essentially gives the bit encoding of natural numbers.
\begin{thm}\label{theorem:parityexists:dynfologn}
\statementDynFOlogn
\end{thm}

In particular, the queries \parityexistsdeg{k}, for  $k \in \N$, do not induce an arity hierarchy for $\DynFO$. For fixed $k$, essentially already unary auxiliary relations suffice.

\begin{thm}\label{theorem:parityexists:dynfodegk}
\statementDynFOdegk
\end{thm}

In both results, Theorem \ref{theorem:parityexists:dynfologn} and \ref{theorem:parityexists:dynfodegk}, the assumption on the presence of a built-in linear order and the $\BIT$ predicate can be lifted, when, for Theorem \ref{theorem:parityexists:dynfologn}, the degree bound of $\parityexistsdeg{\log n}$ refers to the active domain instead of the whole domain, and, for Theorem \ref{theorem:parityexists:dynfodegk}, when binary auxiliary relations are allowed. See Section \ref{section:first-order} for a more detailed discussion.

Finally, we complement our results by a discussion of how queries expressible in $\FO$ extended by arbitrary modulo quantifiers can be maintained in an extension of $\DynFO$. This observation is based on discussions with Samir Datta, Raghav Kulkarni, and Anish Mukherjee.

\subsubsection*{Outline} After recalling the dynamic descriptive complexity scenario in Section \ref{section:preliminaries}, we prove Theorem \ref{theorem:parityexists:dynpropdegk} in Section \ref{section:quantifier-free}, followed by Theorem \ref{theorem:parityexists:dynfologn} and Theorem \ref{theorem:parityexists:dynfodegk} in Section \ref{section:first-order}. 
Section~\ref{section:quasi} contains the discussion regarding maintaining \parityexists and similar queries in extensions of $\DynFO$.
We conclude in Section \ref{section:conclusion}.

\section{Preliminaries and a short introduction to dynamic complexity}
\label{section:preliminaries}
In this article, we write $[n]$ for the set $\{1, \ldots, n\}$ of natural numbers.

We now shortly recapitulate the dynamic complexity framework as introduced by Patnaik and Immerman \cite{PatnaikI97}, and refer to Reference \cite{SchwentickZ16} for details.  

In this framework, a (relational, finite) structure $\inp$ over some schema $\inpSchema$ can be changed by inserting a tuple into or removing a tuple from a relation of $\inp$. 
A \emph{change} $\alpha = \delta(\tpl a)$ consists of an (abstract) \emph{change operation} $\delta$, which is either $\ins_R$ or $\del_R$ for a relation symbol~$R \in \inpSchema$, and a tuple $\tpl a$ over the domain of $\inp$. The change $\ins_R(\tpl a)$ inserts $\tpl a$ into the relation~$R$ of $\inp$, and $\del_R(\tpl a)$ deletes $\tpl a$ from that relation. 
We denote by $\alpha(\inp)$ the structure that results from applying a change $\alpha$ to the structure $\inp$.

A \emph{dynamic program} $\prog$ stores an input structure $\inp$ over $\inpSchema$ as well as an auxiliary structure $\aux$ over some auxiliary schema $\auxSchema$. 
For each change operation $\delta$ and each auxiliary relation $S \in \auxSchema$, the dynamic program has a first-order update rule that specifies how $S$ is updated after a change. Each such rule is of the form \changeRule{$\delta(\tpl p)$}{$S(\tpl x)$}{$\varphi^S_\delta(\tpl p; \tpl x)$} where the \emph{update formula} $\varphi^S_\delta$ is over the combined schema $\inpSchema \cup \auxSchema$ of $\inp$ and $\aux$. Now, for instance, if a tuple $\tpl a$ is inserted into an input relation $R$, the auxiliary relation $S$ is updated to $\{ \tpl b \mid (\inp, \aux) \models \varphi^S_{\ins_R}(\tpl a; \tpl b)\}$. In the standard scenario, all relations in both $\inp$ and $\aux$ are empty initially. 

An \emph{$m$-ary query} $\query$ on $\schema$-structures, for some schema $\schema$, maps each $\schema$-structure with some domain $D$ to a subset of $D^m$, and commutes with isomorphism. A query $\query$ is \emph{maintained} by $\prog$ if $\aux$ has one distinguished relation $\ans$ which, after each sequence of changes, contains the result of $\query$ for the current input structure $\inp$. 

The class $\DynFO$ contains all queries that can be maintained by first-order update rules. The class $\DynProp$ likewise contains the queries that can be maintained by quantifier-free update rules. 
We say that a query $\query$ is in \emph{$k$-ary} \DynFO, for some number $k \in \N$, if it is in \DynFO  via a dynamic program that uses at most $k$-ary auxiliary relations; and likewise for \DynProp.

Sometimes we allow the update formulas to access built-in relations, as for example a predefined linear order $\leq$ and the $\BIT$ predicate. We then assume that the input provides a binary relation that stores a linear order $\leq$ on the domain, which allows to identify the domain with a prefix of the natural numbers, and a binary relation $\BIT$ that contains a tuple $(i,j)$ if the $j$-th bit in the binary representation of $i$ is $1$. Built-in relations are not changed by update rules. 

For expressibility results we will use the standard scenario of Patnaik and Immerman \cite{PatnaikI97} that uses initial input and auxiliary structures with empty relations. Our inexpressibility results are stated for the more powerful scenario where the auxiliary structure may be initialised arbitrarily. See also Reference \cite{ZeumeS15} for a discussion of these different scenarios.

Already quantifier-free programs are surprisingly expressive, as they can maintain, for instance, all regular languages \cite{GeladeMS12} and the transitive closure of deterministic graphs \cite{Hesse03}. As we have seen in the introduction, also the query \Parity can be maintained by quantifier-free update rules. The following example illustrates a standard technique for maintaining queries with quantifier-free update rules which will also be exploited later. 

\begin{exa}\label{example:singletuple:DynProp:klist} For fixed $k \in \N$, let \myproblem{size-$k$} be the Boolean query that asks whether the size of a unary relation \RelName{U} is equal to $k$, so, whether $|\RelName{U}| = k$ holds.
This query is easily definable in \FO for each $k$. We show here that \myproblem{size-$k$} can be maintained by a $\DynProp$-program $\prog$ with binary auxiliary relations.

The dynamic program we construct uses \emph{$k$-lists}, a slight extension of the list technique introduced by Gelade, Marquardt and Schwentick \cite{GeladeMS12}. The list technique was also used to maintain emptiness of a unary relation \RelName{U} under insertions and deletions of single elements with quantifier-free formulas \cite{ZeumeS15}. To this end, a binary relation \RelName{List} is maintained which encodes a linked list of the elements in \RelName{U} in the order of their insertion. Additionally, two unary relations mark the first and the last element of the list. The key insight is that a quantifier-free formula can figure out whether the relation \RelName{U} becomes empty when an element $a$ is deleted by checking whether $a$ is both the first \emph{and} the last element of the list. 

Let us implement this highlevel idea for a fixed $k \in \N$. Let $\ell \df k+1$. To maintain \myproblem{size-$k$}, the quantifier-free dynamic program $\prog$ stores a list of all elements $u \in \RelName{U}$, using a binary relation $\RelName{List}_1$. More precisely, if $u_1, \ldots, u_m$ are the elements in \RelName{U}, then $\RelName{List}_1$ contains the tuples $(u_{i_j},u_{i_{j+1}})$, for $1 \leq j \leq m-1$, where $i_1, \ldots, i_m$ is some permutation of $\{1, \ldots, m\}$. Additionally, the program uses binary relations $\RelName{List}_2, \ldots, \RelName{List}_\ell$ such that $\RelName{List}_i$ describes paths of length $i$ in the linked list $\RelName{List}_1$. For example, if $(u_1,u_2), (u_2,u_3)$ and $(u_3,u_4)$ are tuples in $\RelName{List}_1$, then $(u_1,u_4) \in \RelName{List}_3$.
The list $\RelName{List}_1$  comes with $2\ell$ unary relations $\RelName{First}_1, \ldots, \RelName{First}_\ell, \RelName{Last}_1, \ldots, \RelName{Last}_\ell$ that mark the first and the last $\ell$ elements of the list, as well as with $k+2$ nullary relations $\RelName{Is}_0, \ldots, \RelName{Is}_{k}, \RelName{Is}_{>k}$ that indicate the number of elements in \RelName{U} up to $k$. We call nodes $u$ with $u \in \RelName{First}_i$ or $u \in \RelName{Last}_i$ the $i$-first or the $i$-last element, respectively.

Using these relations, the query can be answered easily: the result is given by $\RelName{Is}_k$. 
We show how to maintain the auxiliary relations under insertions and deletions of single elements, and assume for ease of presentation of the update formulas that if a change $\ins_{\RelName{U}}(u)$ occurs then $u \notin \RelName{U}$ before the change, and a change $\del_{\RelName{U}}(u)$ only happens if $u \in \RelName{U}$ before the change.

\subsubsection*{Insertions of elements}

When an element $u$ is inserted, it needs to be inserted into the list. This element $u$ also becomes the last element of the list (encoded by a tuple $u \in \RelName{Last}_1$), and the $i$-last element $u'$ becomes the $(i+1)$-last one, for $i < \ell$. 
If only $i$ elements are in the list before the change, $u$ becomes the $(i+1)$-first element. The update formulas are as follows: 
\begin{align*}
 \uf{\ins_\RelName{U}}{\RelName{List}_i}{u; x,y} \df & \RelName{List}_i(x,y) \vee \big(\RelName{Last}_i(x) \wedge u = y \big) && \text{\scriptsize for $i \in \{1, \ldots, \ell\}$}  \\ 
 \uf{\ins_\RelName{U}}{\RelName{Last}_1}{u; x} \df &  u = x \allowdisplaybreaks && \\ 
 \uf{\ins_\RelName{U}}{\RelName{Last}_i}{u; x} \df & \RelName{Last}_{i-1}(x)  && \text{\scriptsize for $i \in \{2, \ldots, \ell\}$} \allowdisplaybreaks \\ 
 \uf{\ins_\RelName{U}}{\RelName{First}_i}{u; x} \df & \RelName{First}_i(x) \vee \big( u = x \wedge \RelName{Is}_{i-1} \big)  && \text{\scriptsize for $i \in \{1, \ldots, \ell\}$} \allowdisplaybreaks \\ 
  \uf{\ins_\RelName{U}}{\RelName{Is}_0}{u} \df & \bot  && \allowdisplaybreaks \\ 
  \uf{\ins_\RelName{U}}{\RelName{Is}_i}{u} \df & \RelName{Is}_{i-1}  && \text{\scriptsize for $i \in \{1, \ldots, k\}$} \\   
  \uf{\ins_\RelName{U}}{\RelName{Is}_{>k}}{u} \df & \RelName{Is}_{k} \vee \RelName{Is}_{>k} &&
\end{align*}

\subsubsection*{Deletions of elements}

When an element $u$ is deleted, the hardest task for quantifier-free update formulas is to determine whether, if the size of $\RelName{U}$ was \emph{at least} $k+1$ before the change, its size is now \emph{exactly} $k$. 
We use that if an element $u$ is the $j$-first and at the same time the $j'$-last element, then the list contains exactly $j+j'-1$ elements. If $u$ is removed from the list, $j+j'-2$ elements remain.
So, using the relations $\RelName{First}_j$ and $\RelName{Last}_{j'}$, the exact number $m$ of elements after the change can be determined, if $m \leq 2\ell-2 = 2k$, and in particular, it can be determined whether this number is $k$.
The relations $\RelName{First}_i$ (and, symmetrically the relations $\RelName{Last}_{i}$) can be maintained using the relations $\RelName{List}_j$: if the $i'$-first element $u$ is removed from the list, then $u'$ becomes the $i$-first element for $i' \leq i \leq \ell$ if $(u,u') \in \RelName{List}_{i-i'+1}$. The update formulas exploit these insights:
\begin{align*}
 \uf{\del_\RelName{U}}{\RelName{List}_i}{u; x,y} \df & \big( u \neq x \wedge \bigwedge_{i' \leq i} \neg \RelName{List}_{i'}(x,u) \wedge  \RelName{List}_i(x,y)\big) && \\ 
    & \quad\quad \vee \bigvee_{\substack{j, j' \\ j+j' = i+1}} \big( \RelName{List}_{j}(x,u) \wedge \RelName{List}_{j'}(u,y) \big)  && \text{\scriptsize for $i \in \{1, \ldots, \ell\}$} \allowdisplaybreaks \\ 
 \uf{\del_\RelName{U}}{\RelName{Last}_i}{u; x} \df & \big( \bigwedge_{i' \leq i} \neg \RelName{Last}_{i'}(u) \wedge \RelName{Last}_{i}(x) \big) && \\
         & \quad\quad \vee  \bigvee_{i' \leq i} \big(\RelName{Last}_{i'}(u) \wedge \RelName{List}_{i-i'+1}(x,u)\big)   && \text{\scriptsize for $i \in \{1, \ldots, \ell\}$} \allowdisplaybreaks \\  
  \uf{\del_\RelName{U}}{\RelName{First}_i}{u; x} \df& \big( \bigwedge_{i' \leq i} \neg \RelName{First}_{i'}(u) \wedge \RelName{First}_{i}(x) \big) && \\
        & \quad\quad \vee \bigvee_{i' \leq i} \big(\RelName{First}_{i'}(u) \wedge \RelName{List}_{i-i'+1}(u,x)\big)   && \text{\scriptsize for $i \in \{1, \ldots, \ell\}$} \allowdisplaybreaks \\            
  \uf{\del_\RelName{U}}{\RelName{Is}_i}{u} \df & \bigvee_{\substack{j, j' \\ j+j'-2 = i}} \big( \RelName{First}_{j}(u) \wedge \RelName{Last}_{j'}(u) \big) && \text{\scriptsize for $i \in \{0, \ldots, k\}$} \allowdisplaybreaks \\   
  \uf{\del_\RelName{U}}{\RelName{Is}_{>k}}{u}  \df &  \RelName{Is}_{>k} \wedge \bigwedge_{\substack{j, j' \\ j+j'-2 = k}} \big( \neg \RelName{First}_{j}(u) \vee \neg \RelName{Last}_{j'}(u) \big) &&
\end{align*}
As all auxiliary relations can be maintained under insertions and deletions of elements, the presented dynamic program witnesses that \myproblem{size-$k$} is in binary \DynProp. \qed
\end{exa}

Later on we will use that the dynamic program from Example~\ref{example:singletuple:DynProp:klist} can straightforwardly be extended to maintain the set of nodes in a graph with degree $k$, for each fixed $k \in \N$: for each node $v$ of the graph, the dynamic program maintains whether the set of nodes adjacent to $v$ has size $k$. Towards this end, the program maintains one instance of the auxiliary relations from Example~\ref{example:singletuple:DynProp:klist} for each node $v$. This is realised by increasing the arity of every auxiliary relation by one, and the additional dimension is used to indicate the node to which a tuple belongs. Accordingly, this dynamic program uses at most ternary auxiliary relations.

\section{ParityExists and quantifier-free updates}
\label{section:quantifier-free}

In this section we start our examination of the \parityexists query in the context of quantifier-free update rules. Let us first formalise the query. It is evaluated over \emph{coloured graphs}, that is, directed graphs $(V, E)$ with an additional unary relation $R$ that encodes a set of (red-)\emph{coloured} nodes.\footnote{We note that the additional relation $R$ is for convenience of exposition. All our results are also valid for pure graphs: instead of using the relation $R$ one could consider a node $v$ coloured if it has a self-loop~$(v,v) \in E$.} A node $w$ of such a graph is said to be \emph{covered} if there is a coloured node $v \in R$ with $(v,w) \in E$. The query \parityexists asks, given a coloured graph, whether the number of covered nodes is odd.

As stated in the introduction, \parityexists cannot be maintained with quantifier-free update rules. A closer examination reveals a close connection between a variant of this query and the arity structure of $\DynProp$. Let $k$ be a natural number. The variant \parityexistsdeg{k} of \parityexists asks whether the number of covered nodes that additionally have in-degree at most $k$ is odd. Note that \parityexistsdeg{k} is a query on general coloured graphs, not only on graphs with bounded degree.

\begin{customthm}{\ref{theorem:parityexists:dynpropdegk}}
$\parityexistsdeg{k}$ can be maintained in \DynProp with auxiliary relations of arity $k$, for any $k \geq 3$, but not with auxiliary relations of arity $k-1$, even on graphs with in-degree at most $k$.
\end{customthm}

We repeat two immediate consequences which have already been stated in the introduction.

\begin{customthm}{\ref{theorem:parityexists:dynprop}}
      $\parityexists \not\in \DynProp$. 
\end{customthm}
  
 \begin{customcor}{\ref{corollary:arityhierarchy}}
    \DynProp has a strict arity hierarchy for Boolean graph queries. 
 \end{customcor}  
\begin{proof}
For every $k \geq 1$ we give a Boolean graph query that can be maintained using $k$-ary auxiliary relations, but not with $(k-1)$-ary relations.
 
For $k \geq 3$, we choose the query $\parityexistsdeg{k}$ which satisfies the conditions by Theorem \ref{theorem:parityexists:dynpropdegk}.

For $k = 2$, already  \cite[Proposition 4.10]{ZeumeS15} shows that the query \myproblem{s-t-TwoPath} which asks whether there exists a path of length $2$ between two distinguished vertices $s$ and $t$ separates unary $\DynProp$ from binary \DynProp.

For $k = 1$, we consider the Boolean graph query \myproblem{ParityDegreeDiv3} that asks whether the number of nodes whose degree is divisible by $3$ is odd. 
This query can easily be maintained in \DynProp using only unary auxiliary relations. In a nutshell, a dynamic program can maintain for each node~$v$ the degree of $v$ modulo $3$. 
So, it maintains three unary relations $M_0, M_1, M_2$ with the intention that $v \in M_i$ if the degree of $v$ is congruent to $i$ modulo $3$. These relations can easily be updated under edge insertions and deletions. 
Similar as for \Parity, a bit $P$ that gives the parity of $|M_0|$ can easily be maintained.

On the other hand, \myproblem{ParityDegreeDiv3} cannot be maintained in \DynProp using nullary auxiliary relations. 
Suppose towards a contradiction that it can be maintained by a dynamic program $\prog$ that only uses nullary auxiliary relations and consider an input instance that contains the nodes $V= \{u_1, u_2, v_1, v_2, v_3\}$ and the edges~\mbox{$E = \{(u_1, v_1), (u_1,v_2),(u_2,v_1)\}$}. No matter the auxiliary database, $\prog$ needs to give the same answer after the changes $\alpha_1 \df \ins_E(u_1,v_3)$ and $\alpha_2 \df \ins_E(u_2,v_3)$, as it cannot distinguish these tuples using quantifier-free first-order formulas. But $\alpha_1$ leads to a yes-instance for \myproblem{ParityDegreeDiv3}, and $\alpha_2$ does not. 
See Figure~\ref{figure:dynprop:corollary} for an illustration.
So, $\prog$ does not maintain \myproblem{ParityDegreeDiv3}.
\begin{figure}[t]
\centering
\begin{tikzpicture}[scale=0.7]
\tikzset{rlabel/.style = {font=\scriptsize}}
\tikzset{llabel/.style = {font=\scriptsize}}

\node[mnode, label={[rlabel]below:$u_1$}] at (0,0) (u1) {};
\node[mnode, label={[rlabel]below:$u_2$}] at (2,0) (u2) {};
\node[mnode, label={[rlabel]above:$v_1$}] at (-0.5,2) (v1) {};
\node[mnode, label={[rlabel]above:$v_2$}] at (1,2) (v2) {};
\node[mnode, label={[rlabel]above:$v_3$}] at (2.5,2) (v3) {};

 \path[dEdge] (u1) edge (v1) edge (v2);
 \path[dEdge] (u2) edge (v1);
 \path[dEdge, dashed] (u1) edge (v3);
 \path[dEdge, dotted] (u2) edge (v3);
\end{tikzpicture} 
\caption{The graph used in the proof of Corollary~\ref{corollary:arityhierarchy}. If the dashed edge is included, the graph is a positive instance of \myproblem{ParityDegreeDiv3}. If the dotted edge is included, it is a negative instance.}\label{figure:dynprop:corollary}
\end{figure}
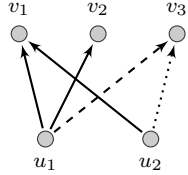
\end{proof}

The rest of this section is devoted to the proof of Theorem \ref{theorem:parityexists:dynpropdegk}. First, in Subsection~\ref{subsec:dynprop:upperbound}, we show  that \parityexistsdeg{k} can be maintained with $k$-ary auxiliary relations, for~$k \geq 3$. For this, we employ the list technique introduced in Example \ref{example:singletuple:DynProp:klist}.  Afterwards, in Subsection \ref{subsec:dynprop:lowerbound}, we prove that auxiliary relations of arity $k-1$ do not suffice. This proof relies on a known tool for proving lower bounds for $\DynProp$ that exploits upper and lower bounds for Ramsey numbers \cite{Zeume17}.
  
\subsection{\texorpdfstring{Maintaining \parityexistsdeg{k}}{Maintaining ParityExists-deg≤k}} \label{subsec:dynprop:upperbound}

We start by proving that $\parityexistsdeg{k}$ can be maintained in \DynProp using $k$-ary auxiliary relations. In Subsection \ref{subsec:dynprop:lowerbound} we show that this arity is optimal.

\begin{prop}\label{prop:dynprop:upperbound}
For every $k \geq 3$, $\parityexistsdeg{k}$ is in $k$-ary \DynProp.
\end{prop}
\begin{proof}
Let $k \geq 3$ be some fixed natural number.
We show how a \DynProp-program $\prog$ can maintain $\parityexistsdeg{k}$ using at most $k$-ary auxiliary relations.

The idea is as follows. 
Whenever a formerly uncoloured node $v$ gets coloured, a certain number $c(v)$ of nodes become covered: $v$ has edges to all these nodes, but no other coloured node has.  Because the number $c(v)$ can be arbitrary, the program $\prog$ necessarily has to store for each uncoloured node $v$ the parity of $c(v)$ to update the query result. But this is not sufficient.  Suppose that another node $v'$ is coloured by a change and that, as a result, a number $c(v')$ of nodes become covered, because they have an edge from $v'$ and so far no incoming edge from another coloured neighbour. Some of these nodes, say, $c(v, v')$ many, also have an incoming edge from $v$.
Of course these nodes do not \emph{become} covered any more when afterwards $v$ is coloured, because they \emph{are} already covered.
So, whenever a node $v'$ gets coloured, the program $\prog$ needs to update the (parity of the) number $c(v)$, based on the (parity of the) number $c(v,v')$. In turn, the (parity of the) latter number needs to be updated whenever another node $v''$ is coloured, using the (parity of the) analogously defined number $c(v,v',v'')$, and so on.

It seems that this reasoning does not lead to a construction idea for a dynamic program, as information for more and more nodes needs to be stored, but observe that only those covered nodes are relevant for the query that have in-degree at most $k$. 
So, a number $c(v_1, \ldots, v_k)$ does not need to be updated when some other node $v_{k+1}$ gets coloured, because no relevant node has edges from all nodes $v_1, \ldots, v_{k+1}$. 

We now present the construction in more detail.
A node $w$ is called \emph{active} if its in-degree $\indeg(w)$ is at most $k$.
Let $A = \{a_1, \ldots, a_\ell\}$ be a set of coloured nodes and let $B = \{b_1, \ldots, b_m\}$ be a set of uncoloured nodes, with $\ell + m \leq k$.
By $\Ncunc_G(A,B)$ we denote the set of active nodes $w$ of the coloured graph $G$ whose coloured \mbox{(in-)}neighbours are exactly the nodes in $A$ and that have (possibly amongst others) the nodes in $B$ as uncoloured \mbox{(in-)}neighbours. 
So, $w \in \Ncunc_G(A,B)$ if 
\begin{enumerate}[(1)]
\item $\indeg(w) \leq k$,
\item $(v,w) \in E$ for all $v \in A \cup B$, and
\item  there is no edge $(v',w) \in E$ from a coloured node $v' \in R$ with $v' \notin A$.
\end{enumerate}
We omit the subscript $G$ and just write $\Ncunc(A,B)$ if the graph $G$ is clear from the context. See Figure~\ref{figure:dynprop:upper} for an example.
The dynamic program $\prog$ maintains the parity of $|\Ncunc_G(A,B)|$ for all such sets $A, B$.

\begin{figure}[t]
\centering
\begin{tikzpicture}[xscale=0.6, yscale=0.6]
\tikzset{rlabel/.style = {font=\scriptsize}}
\tikzset{llabel/.style = {font=\scriptsize}}

\draw [mline, thin] (3.5,-0.45) rectangle (7.0,2);
\draw [mline, thin] (7.5,-0.45) rectangle (9.0,2);
\draw [mline, thin] (2.25,-2.4) rectangle (9.75,-5);

\node at (4,1.5) {\footnotesize $A$};
\node at (8,1.5) {\footnotesize $B$};
\node at (4.55,-4.45) {\footnotesize $\Ncunc(A,B)$};

\pgfmathsetmacro{\rx}{-3} 
\pgfmathsetmacro{\ry}{0} 
\pgfmathsetmacro{\dist}{3}
\pgfmathsetmacro{\lx}{0} 
\pgfmathsetmacro{\ly}{0.25} 
\pgfmathsetmacro{\ldist}{2}
\node[mnode, label={[llabel]above:$v_1$}] at (\ly+0*\ldist,\lx) (1) {};
\node[mnode, label={[llabel]above:$v_2$}] at (\ly+1*\ldist,\lx) (2) {};
\node[mnode, fill=red, label={[llabel]above:$v_3$}] at (\ly+2*\ldist,\lx) (3) {};
\node[mnode, fill=red, label={[llabel]above:$v_4$}] at (\ly+3*\ldist,\lx) (4) {};
\node[mnode, label={[llabel]above:$v_5$}] at (\ly+4*\ldist,\lx) (5) {};
\node[mnode, label={[llabel]above:$v_6$}] at (\ly+5*\ldist,\lx) (6) {};
\node[mnode, fill=red, label={[llabel]above:$v_7$}] at (\ly+6*\ldist,\lx) (7) {};
\node[mnode, label={[rlabel]below:$w_1$}] at (\ry+0*\dist, \rx) (m1) {};
\node[mnode, label={[rlabel]below:$w_2$}] at (\ry+1*\dist, \rx) (m2) {};
\node[mnode, label={[rlabel]below:$w_3$}] at (\ry+2*\dist, \rx) (m3) {};
\node[mnode, label={[rlabel]below:$w_4$}] at (\ry+3*\dist, \rx) (m4) {};
\node[mnode, label={[rlabel]below:$w_5$}] at (\ry+4*\dist, \rx) (m5) {};

 \path[dEdge] (1) edge (m1);
 \path[dEdge] (3) edge (m1) edge (m2) edge (m3) edge (m4) edge (m5);
 \path[dEdge] (2) edge (m2);
 \path[dEdge] (4) edge (m1) edge (m2) edge (m3) edge (m4) edge (m5);
 \path[dEdge] (6) edge (m3);
 \path[dEdge] (5) edge (m2) edge (m3) edge (m4) edge (m5);
 \path[dEdge] (7) edge (m5);
\end{tikzpicture} 
\caption{An illustration of the notation used in the proof of Proposition~\ref{prop:dynprop:upperbound}. The set $\Ncunc(A,B)$ does not include $w_1$, as there is no edge $(v_5,w_1)$, and it does not include $w_5$, as there is an edge $(v_7,w_5)$ for a coloured node $v_7 \not\in A$.}\label{figure:dynprop:upper}
\end{figure}
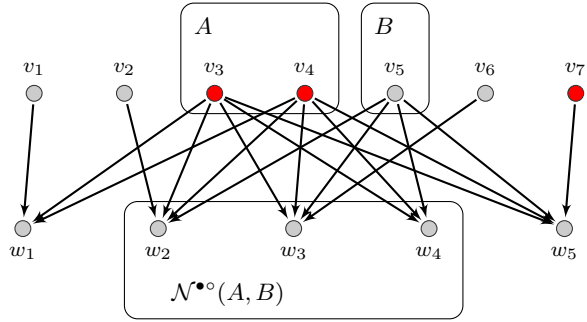

Whenever a change $\alpha = \ins_R(v)$ colours a node $v$ of $G$, the update is as follows. 
We distinguish the three cases 
\begin{enumerate}[(1)]
\item  $v \in A$, 
\item $v \in B$ and 
\item $v \notin A \cup B$. 
\end{enumerate}
In case (1), the set $\Ncunc_{\alpha(G)}(A,B)$ equals the set $\Ncunc_{G}(A \setminus \{v\},B \cup \{v\})$, and the existing auxiliary information can be copied.
In case (2), actually $\Ncunc_{\alpha(G)}(A,B) = \emptyset$, as $B$ contains a coloured node. The parity of the cardinality $0$ of $\emptyset$ is even.
For case (3) we distinguish two further cases. If $|A \cup B| = k$, no active node $w$ can have incoming edges from every node in $A \cup B \cup \{v\}$ as $w$ has in-degree at most $k$, so $\Ncunc_{\alpha(G)}(A,B) = \Ncunc_{G}(A,B)$ and the existing auxiliary information is taken over. 
If $|A \cup  B| < k$, then $\Ncunc_{\alpha(G)}(A,B) = \Ncunc_{G}(A,B) \setminus \Ncunc_{G}(A,B \cup \{v\})$ and $\prog$ can combine the existing auxiliary information.

When a change $\alpha = \del_R(v)$ uncolours a node $v$ of $G$, the necessary updates are symmetrical. 
The case $v \in A$ is similar to case (2) above: $\Ncunc_{\alpha(G)}(A,B) = \emptyset$, because $A$ contains an uncoloured node.
The case $v \in B$ is handled similarly as case (1) above, as we have $\Ncunc_{\alpha(G)}(A,B) = \Ncunc_{G}(A \cup \{v\},B \setminus \{v\})$.
The third case $v \notin A \cup B$ is treated analogously as case (3) above, but in the sub-case $|A \cup  B| < k$ we have that
$\Ncunc_{\alpha(G)}(A,B) = \Ncunc_{G}(A,B) \cup \Ncunc_{G}(A \cup \{v\},B)$.

Edge insertions and deletions are conceptionally easy to handle, as they change the sets $\Ncunc(A,B)$ by at most one element. Given all nodes of $A$ and $B$ and the endpoints of the changed edge as parameters, quantifier-free formulas can easily determine whether this is the case for specific sets $A, B$.

We now present $\prog$ formally.
For every $\ell \leq k+1$ the program maintains unary relations $N_\ell$ and $N_\ell^\bullet$ with the intended meaning that for a node $w$ it holds $w \in N_\ell$ if $\indeg(w) = \ell$ and $w \in N_\ell^\bullet$ if $w$ has exactly $\ell$ coloured in-neighbours.
These relations can be maintained as explained in Example~\ref{example:singletuple:DynProp:klist} and the subsequent remark, requiring some additional, ternary auxiliary relations.
We also use a relation $\RelName{Active} \df N_1 \cup \cdots \cup N_k$ that contains all active nodes with at least one edge.

For every $\ell, m \geq 0$ with $1 \leq \ell + m \leq k$ the programs maintains $(\ell+m)$-ary auxiliary relations $P_{\ell,m}$ with the intended meaning that a tuple $(a_1,\ldots,a_\ell,b_1,\ldots,b_m)$ is contained in $P_{\ell,m}$ if and only if 
\begin{itemize}
 \item the nodes $a_1,\ldots,a_\ell,b_1,\ldots,b_m$ are pairwise distinct,
 \item $a_i \in R$ and $b_j \notin R$ for $i \in [\ell], j \in [m]$, and
 \item the set $\Ncunc(A,B)$ has an odd number of elements, where $A = \{a_1,\ldots,a_\ell\}$ and $B = \{b_1,\ldots,b_m\}$.
\end{itemize}
The following formula $\theta_{\ell,m}$ checks the first two conditions:
 \[\theta_{\ell,m}(x_1,\ldots, x_\ell, y_1, \ldots, y_m) \df \bigwedge_{i \neq j \in [\ell]} x_i \neq x_j \wedge \bigwedge_{i \neq j \in [m]} y_i \neq y_j \wedge \bigwedge_{i \in [\ell]} R(x_i) \wedge \bigwedge_{i \in [m]} \neg R(y_i)\]
Of course, $\prog$ also maintains the Boolean query relation \ans.

We now describe the update formulas of $\prog$ for the relations $P_{\ell,m}$ and $\ans$, assuming that each change actually alters the input graph, so, for example, no changes $\ins_E(v,w)$ occur such that the edge $(v,w)$ already exists.

Let $\varphi \xor \psi \df (\varphi \wedge \neg \psi) \vee (\neg \varphi \wedge \psi)$ denote the Boolean exclusive-or connector.

\subsubsection*{Colouring a node $v$}

A change $\ins_R(v)$ increases the total number of active, covered nodes by the number of active nodes that have so far no coloured in-neighbour, but an edge from $v$. That is, this number is increased by $|\Ncunc(\emptyset,\{v\})|$. 
The update formula for $\ans$ is therefore
\[\uf{\ins_R}{\ans}{v}  \df \ans \xor P_{0,1}(v).\]

We only spell out the more interesting update formulas for the relations $P_{\ell,m}$, for different values of $\ell, m$. These formulas list the conditions for tuples $\tpl a = a_1, \ldots, a_\ell$ and $\tpl b = b_1, \ldots, b_m$ that $\Ncunc(\{a_1, \ldots, a_\ell\},\{b_1, \ldots, b_m\})$ is of odd size after a change.
The other update formulas are simple variants.
\begin{align*}
 \uf{\ins_R}{P_{\ell,m}}{&v; x_1, \ldots, x_\ell, y_1, \ldots, y_m}  \df && \\
  &\bigvee_{i \in [\ell]} \big(v = x_i \wedge P_{\ell-1,m+1}(x_1, \ldots, x_{i-1}, x_{i+1}, \ldots, x_\ell, \tpl y, v)\big) && \\
  \vee \Big(&\bigwedge_{i \in [\ell]} v \neq x_i \wedge \bigwedge_{i \in [m]} v \neq y_i \wedge \big(P_{\ell,m}(\tpl x, \tpl y) \xor P_{\ell,m+1}(\tpl x, \tpl y, v)\big)\Big) && \text{\scriptsize for $\ell \geq 1, \ell+m < k$} \allowdisplaybreaks \\
 \uf{\ins_R}{P_{\ell,m}}{&v; x_1, \ldots, x_\ell, y_1, \ldots, y_m}  \df && \\
  &\bigvee_{i \in [\ell]} \big(v = x_i \wedge P_{\ell-1,m+1}(x_1, \ldots, x_{i-1}, x_{i+1}, \ldots, x_\ell, \tpl y, v)\big) && \\
  \vee \Big(&\bigwedge_{i \in [\ell]} v \neq x_i \wedge \bigwedge_{i \in [m]} v \neq y_i \wedge P_{\ell,m}(\tpl x, \tpl y) \Big) && \text{\scriptsize for $\ell \geq 1, \ell+m = k$} \\  
\end{align*}

\subsubsection*{Uncolouring a node $v$}

 The update formulas for a change $\del_R(v)$ are analogous to the update formulas for a change $\ins_R(v)$ as seen above.
 Again we only present a subset of the update formulas, the others are again easy variants.
 \begin{align*}
  \uf{\del_R}{\ans}{&v}  \df \ans \xor P_{1,0}(v) && \allowdisplaybreaks \\
  \uf{\del_R}{P_{\ell,m}}{&v; x_1, \ldots, x_\ell, y_1, \ldots, y_m}  \df && \\
   &\bigvee_{i \in [m]} \big(v = y_i \wedge P_{\ell+1,m-1}(\tpl x, v, y_1, \ldots, y_{i-1}, y_{i+1}, \ldots, y_m)\big) && \\
   \vee \Big(&\bigwedge_{i \in [\ell]} v \neq x_i \wedge \bigwedge_{i \in [m]} v \neq y_i \wedge \big(P_{\ell,m}(\tpl x, \tpl y) \xor P_{\ell+1,m}(\tpl x, v, \tpl y)\big)\Big) && \text{\scriptsize for $m \geq 1, \ell+m < k$} \allowdisplaybreaks \\
  \uf{\del_R}{P_{\ell,m}}{&v; x_1, \ldots, x_\ell, y_1, \ldots, y_m}  \df && \\
   &\bigvee_{i \in [m]} \big(v = y_i \wedge P_{\ell+1,m-1}(\tpl x, v, y_1, \ldots, y_{i-1}, y_{i+1}, \ldots, y_m)\big) && \\
   \vee \Big(&\bigwedge_{i \in [\ell]} v \neq x_i \wedge \bigwedge_{i \in [m]} v \neq y_i \wedge P_{\ell,m}(\tpl x, \tpl y)\Big) && \text{\scriptsize for $m \geq 1, \ell+m = k$} \\  
 \end{align*}

\subsubsection*{Inserting an edge $(v,w)$}

When an edge $(v,w)$ is inserted, the number of active, covered nodes can change at most by one.
At first, a covered node $w$ might become inactive. This happens when $w$ had in-degree $k$ before the insertion. 
Or, an active node $w$ becomes covered. This happens if $v$ is coloured and $w$ had no coloured in-neighbour and in-degree at most $k-1$ before the change.
The update formula for $\ans$ is accordingly
\[\uf{\ins_E}{\ans}{v,w} \df \ans \xor \Big(\big(N_k(w) \wedge \bigvee_{i \in [k]} N_i^\bullet(w)\big) \vee \big(R(v) \wedge N_0^\bullet(w) \wedge \bigvee_{i \in [k]} N_{i-1}(w) \big)\Big).\]

The necessary updates for relations $P_{\ell,m}$ are conceptionally very similar.
We list the conditions that characterise whether the membership of $w$ in $\Ncunc(A,B)$ changes, for a set $A = \{x_1, \ldots, x_\ell\}$ of coloured nodes and a set $B = \{y_1, \ldots, y_m\}$ of uncoloured nodes.

\begin{itemize}
 \item Before the change, $w \in \Ncunc(A,B)$ holds, but not afterwards. This is either because $w$ becomes inactive or because the new edge $(v,w)$ connects $w$ with another coloured node $v$. This case is expressed by the formula 
 \[\psi_1 \df \bigwedge_{i \in [\ell]} E(x_i,w) \wedge N_\ell^\bullet(w) \wedge \bigwedge_{i \in [m]} E(y_i,w) \wedge \big(N_k(w) \vee R(v)\big).\]
 \item Before the change, $w \in \Ncunc(A,B)$ does not hold, but it does afterwards. Then $w$ needs to be active and to have an incoming edge from all but one node from $A \cup B$, and $v$ is that one node. Additionally, $w$ has no other coloured in-neighbours.
 The following formulas $\psi_2,\psi_3$ express these conditions for the cases $v \in A$ and $v \in B$, respectively.
 \begin{align*}
  \psi_2 \df &\bigvee_{i \in [\ell]} \big( v = x_i \wedge \bigwedge_{j \in [\ell] \setminus \{i\}} E(x_j,w) \wedge \bigwedge_{j \in [m]} E(y_j,w) \wedge N_{\ell-1}^\bullet(w) \wedge \bigvee_{j \in [k]} N_{j-1}(w)\big) \\
  \psi_3 \df &\bigvee_{i \in [m]} \big( v = y_i \wedge \bigwedge_{j \in [m] \setminus \{i\}} E(y_j,w) \wedge \bigwedge_{j \in [\ell]} E(x_j,w) \wedge N_{\ell}^\bullet(w) \wedge \bigvee_{j \in [k]} N_{j-1}(w)\big)
 \end{align*}
\end{itemize}

The update formula for $P_{\ell,m}$ is then
\[\uf{\ins_E}{P_{\ell,m}}{v, w; x_1, \ldots, x_\ell, y_1, \ldots, y_m}  \df \theta_{\ell,m}(\tpl x, \tpl y) \wedge \big(P_{\ell,m}(\tpl x, \tpl y) \xor (\psi_1 \vee \psi_2 \vee \psi_3) \big).  \]

\subsubsection*{Deleting an edge $(v,w)$}

The ideas to construct the update formulas for changes $\del_E(v,w)$ are symmetrical to the constructions for changes $\ins_E(v,w)$.
When an edge $(v,w)$ is deleted, the node $w$ becomes active if its in-degree before the change was $k+1$. It is (still) covered, and then is a new active and covered node, if it has coloured in-neighbours other than $v$. This is the case if $w$ has at least two coloured in-neighbours before the change, or if it has at least one coloured in-neighbour and $v$ is not coloured. 

On the other hand, if $v$ was the only coloured in-neighbour of an active node $w$, this node is not covered any more. 
The update formula for the query relation $\ans$ is therefore 
 \begin{align*}
 \uf{\del_E}{\ans}{v,w} \df \ans \xor \Big(&\big(N_{k+1}(w) \wedge \bigvee_{i \in [k+1]} N_i^\bullet(w) \wedge (\neg R(v) \vee \neg N_1^\bullet(w))\big) \\
  & \vee \big(R(v) \wedge N_1^\bullet(w) \wedge \bigvee_{i \in [k]} N_{i}(w) \big)\Big)
 \end{align*}
 
 Regarding the update of relations $P_{\ell,m}$, we distinguish the same cases as for insertions $\ins_E(v,w)$ for a set $A = \{x_1, \ldots, x_\ell\}$ of coloured nodes and a set $B = \{y_1, \ldots, y_m\}$ of uncoloured nodes.
 
 \begin{itemize}
 \item Before the change, $w \in \Ncunc(A,B)$ holds, but not afterwards. That means the active node $w$ has incoming edges from all nodes from $A \cup B$, has no coloured in-neighbours apart from the nodes in $A$, and $v \in A \cup B$. This is expressed by the formula 
  \begin{align*}
   \psi'_1 \df & \bigwedge_{i \in [\ell]} E(x_i,w) \wedge N_\ell^\bullet(w) \wedge \bigwedge_{i \in [m]} E(y_i,w) \wedge \bigvee_{i \in [k]} N_i(w) \\
    & \qquad \wedge \big( \bigvee_{i \in [\ell]} x_i = v \vee \bigvee_{i \in [m]} y_i = v\big).
  \end{align*}
  
  \item Before the change, $w \in \Ncunc(A,B)$ does not hold, but it does afterwards. Then $w$ already is and stays connected to all nodes from $A \cup B$, and either have degree $k+1$ and become active and/or loose an additional coloured in-neighbour $v$.
 The following formula $\psi'_2$ lists the possible combinations.
  \begin{align*}
   \psi'_2 \df &\bigwedge_{i \in [\ell]} (v \neq x_i \wedge E(x_i,w)) \wedge \bigwedge_{i \in [m]} (v \neq y_i \wedge E(y_i,w)) \\
  & \qquad \wedge \Big(\big(N_{k+1}(w) \wedge \neg R(v) \wedge N_{\ell}^\bullet(w)\big) \\
  & \qquad \quad \vee \big( N_{k+1}(w) \wedge R(v) \wedge N_{\ell+1}^\bullet(w)\big) \\
  & \qquad \quad \vee \big( \bigvee_{i \in [k]}N_{i}(w) \wedge R(v) \wedge N_{\ell+1}^\bullet(w)\big)\Big)
  \end{align*}
 \end{itemize}

 Finally, the update formula for $P_{\ell,m}$ is
 \[\uf{\del_E}{P_{\ell,m}}{v, w; x_1, \ldots, x_\ell, y_1, \ldots, y_m}  \df \theta_{\ell,m}(\tpl x, \tpl y) \wedge \big(P_{\ell,m}(\tpl x, \tpl y) \xor (\psi'_1 \vee \psi'_2) \big).  \qedhere\] 
\end{proof}

Our proof does not go through for $k < 3$, as we use ternary auxiliary relations to maintain whether a node has degree at most $k$, see Example~\ref{example:singletuple:DynProp:klist} and the subsequent remark. 
In fact, this cannot be circumvented, as formalised by the next proposition.

\begin{prop}\label{prop:dynprop:upperbound:min3}
For $k \in \{1,2\}$, $\parityexistsdeg{k}$ is not in binary \DynProp, even with arbitrary initialisation.
\end{prop}

The proof relies on a lower bound result by Zeume and Schwentick \cite{ZeumeS15}. They show that one cannot maintain in binary \DynProp, not even with arbitrary initialisation, whether there is a directed path from some distinguished node $s$ to some distinguished node $t$ in a $2$-layered graph $G$ \cite[Theorem~4.7]{ZeumeS15}.
A graph $G = (V,E,s,t)$ with distinguished nodes $s,t \in V$ is \emph{$2$-layered}, if its set $V$ of nodes can be partitioned into sets $V = \{s,t\} \cup A \cup B$, such that edges go either from $s$ to some node in $A$, from some node in $A$ to some node in $B$, or from some node in $B$ to $t$. So, the edge set $E$ is a subset of $(\{s\} \times A) \cup (A \times B) \cup (B \times \{t\})$.

We prove Proposition~\ref{prop:dynprop:upperbound:min3} using a reduction from (a special case of) this query.

\begin{proof}
We first show the result for $k=1$. 
The proof of \cite[Theorem~4.7]{ZeumeS15} shows that there is no dynamic program with quantifier-free update rules and binary auxiliary relations that can maintain $s$-$t$-reachability in $2$-layered graphs, not even if the auxiliary relations may be initialised arbitrarily.
The proof actually shows that such dynamic programs cannot even maintain the query if
\begin{itemize}
 \item the initial graph may be any $2$-layered graph $G = (V,E,s,t)$ with some node set $V = \{s,t\} \cup A \cup B$ that satisfies the following conditions: 
  \begin{itemize}
   \item there is no edge from $s$ to any other node, and
   \item all nodes from $B$ have an edge to $t$,
  \end{itemize}
  \item the auxiliary relations are initialised arbitrarily,
  \item the changes consist of a sequence of deletions of edges from $A \times B$ followed by the insertion of a single edge from $s$ to some node in $A$.
\end{itemize}

We now assume, towards a contradiction, that there is a dynamic program $\prog'$ that witnesses that $\parityexistsdeg{1}$ is in binary \DynProp. We show that from $\prog'$ we can construct a dynamic program $\prog$ with quantifier-free update rules and binary auxiliary relations that can maintain $s$-$t$-reachability for $2$-layered graphs with the restrictions noted above, contradicting \cite[Theorem~4.7]{ZeumeS15}.

Let $G = (V,E,s,t)$ be some $2$-layered graph with node set $V = \{s,t\} \cup A \cup B$, such that $(b,t) \in E$ for every $b \in B$ and $(s,v) \not\in E$ for every $v \in V$.
Let $G' = (V,E',R)$ be the coloured graph with the same node set $V$ as $G$, edge set $E' \df \{(v,u) \mid (u,v) \in E\}$, and $R = \emptyset$. Let $\aux'$ be the auxiliary relations that $\prog'$ assigns to $G'$ starting from an initially empty graph and arbitrarily initialised auxiliary relations when the edges $E'$ are inserted in some arbitrary order.

We now explain how a dynamic program $\prog$ can maintain $s$-$t$-reachability for $G$ under deletions of edges from $A \times B$ followed by the insertion of a single edge from $s$ to some node in $A$, starting from the initial auxiliary relations $\aux'$. 
The basic idea is that changes to $G$ are translated to changes to $G'$ such that 
\begin{itemize}
\item if the graph obtained from $G$ has an $s$-$t$-path then the graph obtained from $G'$ has no covered node with in-degree at most $1$, and
\item if the graph obtained from $G$ has no $s$-$t$-path then the graph obtained from $G'$ has exactly one covered node with in-degree at most $1$.
\end{itemize}

The dynamic program $\prog$ proceeds as follows.
If an edge $(u,v) \in E$ with $u \in A$ and  $v \in B$ is deleted, then $\prog$ simulates $\prog'$ for the deletion of the edge $(v,u) \in E'$.
If an edge $(s,a)$ is inserted into $E$, then $\prog$ simulates $\prog'$ for the insertion of $(s,a)$ into $E'$ followed by the insertion of $s$ into $R$. Then $\prog$ gives the query answer ``true'' if and only if the answer of~$\prog'$ is ``false''.
All this is clearly expressible by quantifier-free update formulas.

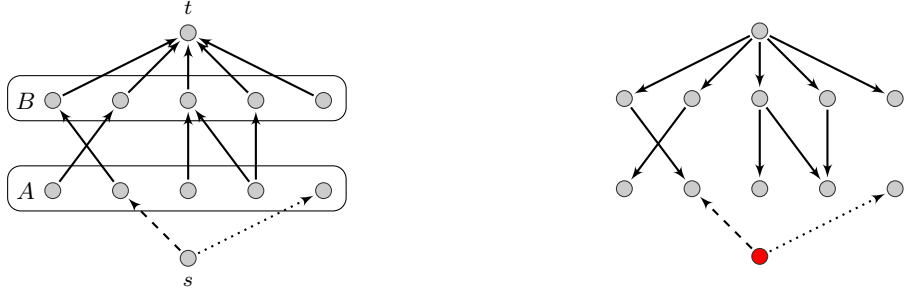
\begin{figure}[t]
\centering
\begin{subfigure}{.45\linewidth}
\centering
\begin{tikzpicture}[scale=0.6]
\tikzset{rlabel/.style = {font=\scriptsize}}
\tikzset{llabel/.style = {font=\scriptsize}}

\draw [mline, thin] (-1,-0.45) rectangle (6.5,.55);
\draw [mline, thin] (-1,-2.45) rectangle (6.5,-1.45);

\node at (-.6,0) {\footnotesize $B$};
\node at (-.6,-2) {\footnotesize $A$};

\pgfmathsetmacro{\rx}{-2} 
\pgfmathsetmacro{\ry}{0} 
\pgfmathsetmacro{\dist}{1.5}
\pgfmathsetmacro{\lx}{0} 
\pgfmathsetmacro{\ly}{0} 
\pgfmathsetmacro{\ldist}{1.5}
\node[mnode] at (\ly+0*\ldist,\lx) (1) {};
\node[mnode] at (\ly+1*\ldist,\lx) (2) {};
\node[mnode] at (\ly+2*\ldist,\lx) (3) {};
\node[mnode] at (\ly+3*\ldist,\lx) (4) {};
\node[mnode] at (\ly+4*\ldist,\lx) (5) {};
\node[mnode] at (\ry+0*\dist, \rx) (m1) {};
\node[mnode] at (\ry+1*\dist, \rx) (m2) {};
\node[mnode] at (\ry+2*\dist, \rx) (m3) {};
\node[mnode] at (\ry+3*\dist, \rx) (m4) {};
\node[mnode] at (\ry+4*\dist, \rx) (m5) {};
\node[mnode, label={[rlabel]below:$s$}] at (3, -3.5) (s) {};
\node[mnode, label={[rlabel]above:$t$}] at (3, 1.5) (t) {};

 \path[dEdge] (1) edge (t);
 \path[dEdge] (3) edge (t);
 \path[dEdge] (2) edge (t);
 \path[dEdge] (4) edge (t);
 \path[dEdge] (5) edge (t);
 \path[dEdge] (m1) edge (2);
 \path[dEdge] (m2) edge (1);
 \path[dEdge] (m3) edge (3);
 \path[dEdge] (m4) edge (3) edge (4);
 \path[dEdge, dashed] (s) edge (m2);
 \path[dEdge, dotted] (s) edge (m5);
\end{tikzpicture} 
\caption{A $2$-layered graph $G_u$. If the dashed edge is included, the graph contains an $s$-$t$-path. If the dotted edge is included, the graph has no such path.}
\end{subfigure} \qquad
\begin{subfigure}{.45\linewidth}
\centering
\begin{tikzpicture}[scale=0.6]
\tikzset{rlabel/.style = {font=\scriptsize}}
\tikzset{llabel/.style = {font=\scriptsize}}

\pgfmathsetmacro{\rx}{-2} 
\pgfmathsetmacro{\ry}{0} 
\pgfmathsetmacro{\dist}{1.5}
\pgfmathsetmacro{\lx}{0} 
\pgfmathsetmacro{\ly}{0} 
\pgfmathsetmacro{\ldist}{1.5}
\node[mnode] at (\ly+0*\ldist,\lx) (1) {};
\node[mnode] at (\ly+1*\ldist,\lx) (2) {};
\node[mnode] at (\ly+2*\ldist,\lx) (3) {};
\node[mnode] at (\ly+3*\ldist,\lx) (4) {};
\node[mnode] at (\ly+4*\ldist,\lx) (5) {};
\node[mnode] at (\ry+0*\dist, \rx) (m1) {};
\node[mnode] at (\ry+1*\dist, \rx) (m2) {};
\node[mnode] at (\ry+2*\dist, \rx) (m3) {};
\node[mnode] at (\ry+3*\dist, \rx) (m4) {};
\node[mnode] at (\ry+4*\dist, \rx) (m5) {};
\node[mnode, fill=red, label={[rlabel]below:$ $}] at (3, -3.5) (s) {};
\node[mnode, label={[rlabel]above:$ $}] at (3, 1.5) (t) {};

 \path[dEdge] (t) edge (1);
 \path[dEdge] (t) edge (2);
 \path[dEdge] (t) edge (3);
 \path[dEdge] (t) edge (4);
 \path[dEdge] (t) edge (5);
 \path[dEdge] (2) edge (m1);
 \path[dEdge] (1) edge (m2);
 \path[dEdge] (3) edge (m3) edge (m4);
 \path[dEdge] (4) edge (m4);
 \path[dEdge, dashed] (s) edge (m2);
 \path[dEdge, dotted] (s) edge (m5);
\end{tikzpicture} 
\caption{A coloured graph $G'_u$. If the dashed edge is included, it is a negative instance of $\parityexistsdeg{1}$. If the dotted edge is included, it is a positive instance.}
\end{subfigure}
\caption{An example for the construction in the proof of Proposition~\ref{prop:dynprop:upperbound:min3}.}\label{figure:dynprop:upperbound:min3}
\end{figure}

Let $G_u$ be the $2$-layered graph that results from $G$ by applying the changes, and let~$G'_u$ be the coloured graph that results from $G'$ by applying the corresponding changes. An example is depicted in Figure~\ref{figure:dynprop:upperbound:min3}. It remains to show that there is an $s$-$t$-path in $G_u$ if and only if the number of covered nodes in $G'_u$ that have in-degree at most $1$ is not odd.

Suppose that there is an $s$-$t$-path in $G_u$ using edges $(s,a),(a,b),(b,t)$, for some nodes $a \in A$ and $b \in B$. It follows that the edges $(s,a)$ and $(b,a)$ are present in $G'_u$, so, $a$ has in-degree at least $2$. As $a$ is the only neighbour of the only red node $s$, there are $0$ covered nodes in $G'_u$ with in-degree at most $1$, an even number.
Suppose on the other hand that there is no $s$-$t$-path in $G_u$. That means that the only node $a \in A$ such that an edge $(s,a)$ exists in $G_u$ has no edge to any node in $B$, because every node from $B$ has an edge to $t$. Consequently, the only incoming edge of $a$ in $G'_u$ is from $s$, which is a red node. So, there is exactly one covered node in $G'_u$ with in-degree at most $1$, an odd number. This concludes the case $k=1$.

The case of $k=2$ can be proven along the same lines, with the only adjustment that in the graph $G'$ we add edges from $t$ to every node in $A \cup B$, increasing the in-degree of every node from this set by $1$.
\end{proof}

\subsection{\texorpdfstring{Inexpressibility results for \parityexistsdeg{k}}{Inexpressibility results for ParityExists-deg≤k}} \label{subsec:dynprop:lowerbound}

In this subsection we prove that $k$-ary auxiliary relations are not sufficient to maintain $\parityexistsdeg{k+1}$, for every~$k \in \N$.
The proof technique we use, and formalise as Lemma~\ref{lemma:dynprop:lowertechnique}, is a reformulation of the proof technique of Zeume \cite{Zeume17}, which combines proof techniques from dynamic complexity \cite{GeladeMS12, ZeumeS15} with insights regarding upper and lower bounds for Ramsey numbers. We actually use a special case of the formalisation from \cite[Lemma~7.4]{SchwentickVZ18}, which is sufficient for our application.

The technique consists of a sufficient condition under which a Boolean query $\query$ cannot be maintained in \DynProp with at most $k$-ary auxiliary relations. 
The condition basically requires that for each collection $\calB$ of subsets of size $k+1$ of a set $\{1,\ldots,n\}$ of elements that may be changed, for an arbitrary~$n$, there is a structure $\db$ whose domain includes the elements $\{1,\ldots,n\}$ and a sequence $\delta_1(x_1), \ldots, \delta_{k+1}(x_{k+1})$ of changes such that 
\begin{enumerate}[(1)]
\item the elements $1, \ldots, n$ cannot be distinguished by quantifier-free formulas evaluated on $\db$, and 
\item the structure that results from $\db$ by applying the changes $\delta_1(i_1), \ldots, \delta_{k+1}(i_{k+1})$ in that order, for each choice of $k+1$ elements $\{i_1, \ldots i_{k+1}\} \subseteq \{1, \ldots, n\}$, is a positive instance for $\query$ exactly if $\{i_1, \ldots,i_{k+1}\} \in \calB$. 
\end{enumerate}

In the following, we write $(\db, \tpl a) \equiv_0 (\db, \tpl b)$ if $\tpl a$ and $\tpl b$ have the same length and agree on their quantifier-free type in $\db$, that is, $\db \models \psi(\tpl a)$ if and only if $\db \models \psi(\tpl b)$ for all quantifier-free formulas $\psi$. We denote the set of all subsets of size $k$ of a set $A$ by $\kpower[k]{A}$.

\begin{lemC}[\cite{SchwentickVZ18}] \label{lemma:dynprop:lowertechnique}
Let $\query$ be a Boolean query of $\schema$-structures.
Then $\query$ is not in $k$-ary \DynProp, even with arbitrary initialisation, if for each $n \in \N$ and all subsets $\calB \subseteq \kpower[k+1]{[n]}$ there exist
 \begin{itemize}
  \item  a $\schema$-structure $\db$ and a set $P = \{p_1, \ldots,p_n\}$ of distinct elements such that
    \begin{itemize}
    \item $P$ is a subset of the domain of \db,
    \item $(\db,p_{i_1}, \ldots, p_{i_{k+1}}) \equiv_0 (\db, p_{j_1}, \ldots, p_{j_{k+1}})$ for all strictly increasing sequences $i_1, \ldots, i_{k+1}$ and $j_1, \ldots, j_{k+1}$ over $[n]$, and
    \end{itemize}
 \item a sequence $\delta_1(x_1), \ldots, \delta_{k+1}(x_{k+1})$ of changes
\end{itemize}
such that for all strictly increasing sequences $i_1, \ldots, i_{k+1}$ over $[n]$:
\[(\delta_1(p_{i_1}) \circ \ldots \circ \delta_{k+1}(p_{i_{k+1}}))(\db) \in \query \quad \Longleftrightarrow \quad \{i_1, \ldots, i_{k+1}\}  \in \calB.\]
\end{lemC}

With the help of Lemma~\ref{lemma:dynprop:lowertechnique} we can show the desired inexpressibility result.

\begin{prop}\label{prop:dynprop:lowerbound}
For every $k \geq 0$, $\parityexistsdeg{k+1}$ is not in $k$-ary \DynProp, even with arbitrary initialisation. This even holds if the input graph may only have in-degree at most $k+1$.
\end{prop}

In the following, for a graph $G = (V,E)$ and some set $X \subseteq V$ of nodes we write $\NoutE(X)$ for the set $\{v \mid \exists u \in X \colon E(u,v)\}$ of out-neighbours of the nodes in $X$. For singleton sets $X = \{x\}$ we just write $\NoutE(x)$ instead of $\NoutE(\{x\})$. Similarly, we write $\NoutA(X)$ for the set $\{v \mid \forall u \in X \colon E(u,v)\} = \bigcap_{x \in X} \NoutE(x)$ of nodes that are out-neighbours of every node in $X$.

\begin{proof}
Let $k \in \N$ be fixed. We apply Lemma \ref{lemma:dynprop:lowertechnique} to show that $\parityexistsdeg{k+1}$ is not in $k$-ary \DynProp. 

The basic proof idea is simple. Given a collection $\calB \subseteq \kpower[k+1]{[n]}$, we construct a graph $G = (V,E)$ with distinguished nodes $P = \{p_1, \ldots, p_n\} \subseteq V$ such that 
\begin{enumerate}[(1)]
\item each node has in-degree at most $k+1$ and 
\item for each $B \in \kpower[k+1]{[n]}$ the set $\NoutE(\{p_i \mid i \in B\})$ is of odd size if and only if $B \in \calB$. 
\end{enumerate}
Then applying a change sequence $\alpha$ which colours all nodes $p_i$ with $i \in B$ to $G$ results in a positive instance of \parityexistsdeg{k+1} if and only if $B \in \calB$. An invocation of Lemma~\ref{lemma:dynprop:lowertechnique} yields the intended lower bound.

It remains to construct the graph $G$. Let $S$ be the set of all non-empty subsets of $[n]$ of size at most $k+1$.
We choose the node set $V$ of $G$ as the union of $P$ and $S$. Only nodes in $P$ will be coloured, and only nodes from $S$ will be covered.
A first attempt to realise the idea mentioned above might be to consider an edge set $\{(p_i, B) \mid B \in \calB, i \in B\}$: then, having fixed some set $B \in \calB$, the node $B$ becomes covered whenever the nodes $p_i$ with $i \in B$ are coloured.
However, also some nodes $B' \neq B$ will be covered, namely if $B' \cap B \neq \emptyset$, and the number of these nodes influences the query result.
We ensure that the set of nodes $B' \neq B$ that are covered by $\{p_i \mid i \in B\}$ is of even size, so that the parity of $|\NoutE(\{p_i \mid i \in B\})|$ is determined by whether $B \in \calB$ holds. This will be achieved by introducing edges to nodes $\kpower[i]{[n]} \in S$ for $i \leq k$ such that for every subset $P'$ of $P$ of size at most $k$ the number of nodes from $S$ that have an incoming edge from \emph{all} nodes from~$P'$ is even. 
By an inclusion-exclusion argument we conclude that for any set $\hat{P} \in \kpower[k+1]{P}$ the number of nodes from $S$ that have an incoming edge from \emph{some} node of $\hat{P}$, but not from all of them, is even.
It follows that whenever $k+1$ nodes $p_{i_1}, \ldots, p_{i_{k+1}}$ are marked, the number of covered nodes is odd precisely if there is one node in $S$ that has an edge from \emph{all} nodes $p_{i_1}, \ldots, p_{i_{k+1}}$, which is the case exactly if $\{i_1, \ldots, i_{k+1}\} \in \calB$.

We now make this precise. Let $n$ be arbitrary and let $P = \{p_1, \ldots, p_n\}$. 
For a set $X \subseteq [n]$ we write $P_X$ for the set $\{p_i \mid i \in X\}$.

The structure $\db$ we construct consists of a coloured graph $G = (V, E, R)$ with nodes $V \df P \cup S$, where $S \df \kpower[1]{[n]} \cup \cdots \cup \kpower[k+1]{[n]}$, and initially empty set $R \df \emptyset$ of coloured nodes.
The edge set $E$ contains all edges $(p_i, Y)$ such that $i \in Y$ and the set $\{B \in \calB \mid Y \subseteq B\}$ has odd size.
See Figure \ref{figure:dynprop:lowerbound} for an example of this construction.
Note that for each $Y \in \kpower[i]{[n]}$, for $i \in \{1, \ldots, k+1\}$, the degree of $Y$ in $G$ is at most $i$, and therefore also at most $k+1$.
 
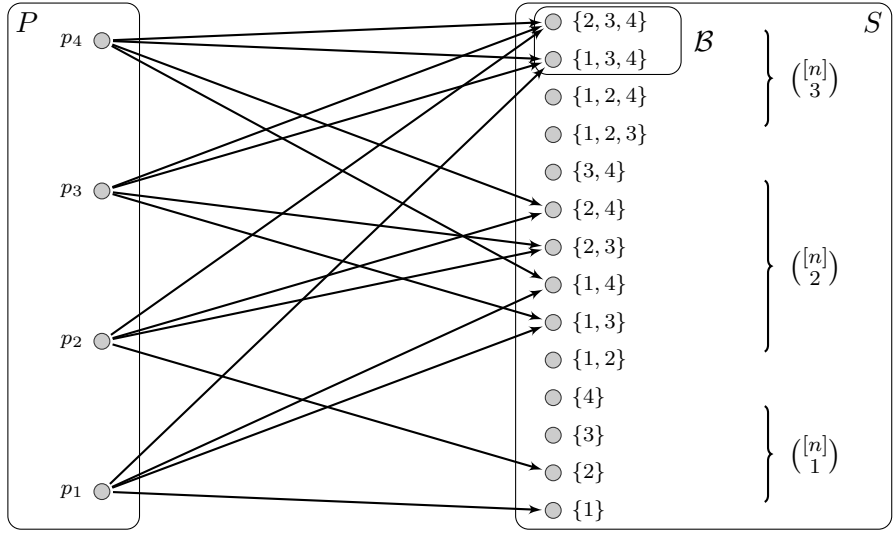
\begin{figure}[t]
\centering
\begin{tikzpicture}
\tikzset{rlabel/.style = {font=\scriptsize}}
\tikzset{llabel/.style = {font=\scriptsize}}

\draw [mline, thin] (-1.25,-0.25) rectangle (0.5,6.75); %
\draw [mline, thin] (5.5,-0.25) rectangle (10.5,6.75); %
\draw [mline, thin] (5.75,5.8) rectangle (7.7,6.7);
 \node at (8,6.25) {$\calB$};
 \node at (-1,6.5) {$P$};
 \node at (10.25,6.5) {$S$};
\pgfmathsetmacro{\rx}{6} 
\pgfmathsetmacro{\ry}{0} 
\pgfmathsetmacro{\dist}{0.5}
\pgfmathsetmacro{\lx}{0} 
\pgfmathsetmacro{\ly}{0.25} 
\pgfmathsetmacro{\ldist}{2}
\node[mnode, label={[llabel]left:$p_1$}] at (\lx,\ly+0*\ldist) (1) {};
\node[mnode, label={[llabel]left:$p_2$}] at (\lx,\ly+1*\ldist) (2) {};
\node[mnode, label={[llabel]left:$p_3$}] at (\lx,\ly+2*\ldist) (3) {};
\node[mnode, label={[llabel]left:$p_4$}] at (\lx,\ly+3*\ldist) (4) {};
\node[mnode, label={[rlabel]right:$\{1\}$}] at (\rx,\ry+0*\dist) (m1) {};
\node[mnode, label={[rlabel]right:$\{2\}$}] at (\rx,\ry+1*\dist) (m2) {};
\node[mnode, label={[rlabel]right:$\{3\}$}] at (\rx,\ry+2*\dist) (m3) {};
\node[mnode, label={[rlabel]right:$\{4\}$}] at (\rx,\ry+3*\dist) (m4) {};
\node[mnode, label={[rlabel]right:$\{1,2\}$}] at (\rx,\ry+4*\dist) (m12) {};
\node[mnode, label={[rlabel]right:$\{1,3\}$}] at (\rx,\ry+5*\dist) (m13) {};
\node[mnode, label={[rlabel]right:$\{1,4\}$}] at (\rx,\ry+6*\dist) (m14) {};
\node[mnode, label={[rlabel]right:$\{2,3\}$}] at (\rx,\ry+7*\dist) (m23) {};
\node[mnode, label={[rlabel]right:$\{2,4\}$}] at (\rx,\ry+8*\dist) (m24) {};
\node[mnode, label={[rlabel]right:$\{3,4\}$}] at (\rx,\ry+9*\dist) (m34) {};
\node[mnode, label={[rlabel]right:$\{1,2,3\}$}] at (\rx,\ry+10*\dist) (m123) {};
\node[mnode, label={[rlabel]right:$\{1,2,4\}$}] at (\rx,\ry+11*\dist) (m124) {};
\node[mnode, label={[rlabel]right:$\{1,3,4\}$}] at (\rx,\ry+12*\dist) (m134) {};
\node[mnode, label={[rlabel]right:$\{2,3,4\}$}] at (\rx,\ry+13*\dist) (m234) {};
\draw[thick, decoration={brace,mirror,raise=80pt},decorate]
  (m123) -- node[right=85pt] {$\kpower[3]{[n]}$} (m234);
\draw[thick, decoration={brace,mirror,raise=80pt},decorate]
  (m12) -- node[right=85pt] {$\kpower[2]{[n]}$} (m34);
\draw[thick, decoration={brace,mirror,raise=80pt},decorate]
  (m1) -- node[right=85pt] {$\kpower[1]{[n]}$} (m4);    

\path[dEdge] (2) edge (m234);
\path[dEdge] (3) edge (m234);
\path[dEdge] (4) edge (m234);

\path[dEdge] (1) edge (m134);
\path[dEdge] (3) edge (m134);
\path[dEdge] (4) edge (m134);

\path[dEdge] (2) edge (m24);
\path[dEdge] (4) edge (m24);
\path[dEdge] (2) edge (m23);
\path[dEdge] (3) edge (m23);
\path[dEdge] (1) edge (m14);
\path[dEdge] (4) edge (m14);
\path[dEdge] (1) edge (m13);
\path[dEdge] (3) edge (m13);
\path[dEdge] (2) edge (m2);
\path[dEdge] (1) edge (m1);

\end{tikzpicture} 
\caption{Example for the construction in the proof of Proposition \ref{prop:dynprop:lowerbound}, with $k = 2$ and $n=4$.}\label{figure:dynprop:lowerbound}
\end{figure}

We now show that for a set $B \in \kpower[k+1]{[n]}$ the cardinality of $\NoutE(P_B)$ is indeed odd if and only if $B \in \calB$.
This follows by an inclusion-exclusion argument. 
For a set $B \in \kpower[k+1]{[n]}$ the set $\NoutE(P_B)$ contains all nodes with an incoming edge from a node in $P_B$. It is therefore equal to the union $\bigcup_{i \in B} \NoutE(p_i)$.  When we sum up the cardinalities of these sets $\NoutE(p_i)$, any node in $\NoutE(P_B)$ with edges to both $p_i$ and $p_j$, for numbers $i,j \in B$, is accounted for twice. 
Continuing this argument, the cardinality of $\NoutE(P_B)$ can be computed as follows:

 \[\big|\NoutE(P_B)\big| = \sum_{i \in B} \big|\NoutA(p_i)\big| 
   - \sum_{\substack{i,j \in B \\ i < j}} \big|\NoutA(\{p_i, p_j\})\big| 
   + \cdots 
   + (-1)^{k} \big|\NoutA(P_B)\big|.\]
For the parity of $|\NoutE(P_B)|$ it therefore holds

 \[\big|\NoutE(P_B)\big| \equiv \sum_{\emptyset \subsetneq X \subseteq B} \big|\NoutA(P_X)\big| \pmod 2.\]
Our claim follows from the observation that $|\NoutA(P_X)|$ is odd if and only if $X \in \calB$, which we will prove next.

Recall that for each $i \leq n$ and $Y \in S$ an edge $(p_i, Y)$ exists precisely if $i \in Y$ and $Y$ is included in an odd number of sets $B' \in \calB$.
This means that $Y \in \NoutA(P_X)$ if and only if $X \subseteq Y$ and $Y$ is included in an odd number of sets $B' \in \calB$.
The parity of $|\NoutA(P_X)|$ is therefore congruent modulo $2$ to the number of pairs $(Y,B')$ such that $X \subseteq Y \subseteq B'$ and $B' \in \calB$ hold: if $Y$ is included in an even number of sets $B' \in \calB$ then $Y \not\in \NoutA(P_X)$, but the parity is not affected by also counting the even number of pairs $(Y,B')$ for this set $Y$. 

For a fixed set $B'$, the number of sets $Y$ with $X \subseteq Y \subseteq B'$ is $2^{|B' \setminus X|}$. We conclude that 

\[ \big|\NoutA(P_X)\big| \equiv \sum_{B' \in \calB, X \subseteq B'} 2^{|B' \setminus X|} \pmod 2, \]
where the right-hand side is odd if and only if there is a $B' \in \calB$ with $X \subseteq B'$ and $B' \setminus X = \emptyset$, so if $X \in \calB$. This proves the claim.

Let $\delta_1(x_1), \ldots, \delta_{k+1}(x_{k+1})$ be the change sequence $\ins_R(x_1), \ldots, \ins_R(x_{k+1})$ that colours the nodes $x_1, \ldots, x_{k+1}$.
Let $B \in \kpower[k+1]{[n]}$ be of the form $\{i_1, \ldots, i_{k+1}\}$ with $i_1 < \cdots < i_{k+1}$. 
The change sequence $\alpha_B \df \delta_1(p_{i_1}) \cdots  \delta_{k+1}(p_{i_{k+1}}) = \ins_R(p_{i_1}), \ldots, \ins_R(p_{i_{k+1}})$ results in a graph where the set of coloured nodes is exactly $P_B$. 
As all nodes in $\NoutE(P_B)$ have degree at most $k+1$ and the set $\NoutE(P_B)$ is of odd size exactly if $B \in \calB$, we have that $\alpha_B(\db)$ is a positive instance of \parityexistsdeg{k+1} if and only if $B \in \calB$.
\end{proof}

\section{ParityExists and first-order updates}
\label{section:first-order}

As discussed in the introduction, the \Parity query can be easily maintained with first-order update rules. So far we have seen that its generalisation $\parityexists$ can only be maintained with quantifier-free update rules if the in-degree of covered nodes is bounded by a constant. Now we show that with full first-order update rules, this query can be maintained if the in-degree is bounded by $\log n$, where $n$ is the number of nodes in the graph. We emphasise that only the in-degree of covered nodes is bounded, while a coloured node $v$ can cover arbitrarily many nodes. If also the out-degree of coloured node is restricted, maintenance in \DynFO becomes trivial\footnote{If the out-degree of a node is upper-bounded by some constant $c$ and the colour of some node $v$ changes, then first-order update rules can determine the number $i \leq c$ of nodes that have an incoming edge from $v$ and no coloured in-neighbour $w \neq v$. The change of the query result is determined by the parity of $i$.}. 

We start by providing a dynamic program with first-order update rules that maintains $\parityexistsdeg{k}$, for a constant $k$, and only uses unary relations apart from a linear order. Thus, in contrast to quantifier-free update rules, this query cannot be used to obtain an arity hierarchy for graph queries for first-order update rules. Afterwards we will exploit the technique used here to maintain $\parityexistsdeg{\log n}$ with binary auxiliary relations.

\begin{customthm}{\ref{theorem:parityexists:dynfodegk}}
\statementDynFOdegk
\end{customthm}

An intuitive reason why quantifier-free dynamic programs for  $\parityexistsdeg{k}$ need auxiliary relations of growing arity is that for checking whether some change, for instance the colouring of a node $v$, is ``relevant'' for some node $w$, it needs to have access to all of $w$'s ``important'' neighbours. Without quantification, the only way to do this is to explicitly list them as elements of the tuple for which the update formula decides whether to include it in the auxiliary relation.

With quantification and a linear order, sets of neighbours can be defined more easily, if the total number of neighbours is bounded by a constant. Let us fix a node $w$ with at most $k$ (in-)neighbours, for some constant $k$. Thanks to the linear order, the neighbours can be distinguished as first, second, \dots, $k$-th neighbour of $w$, and any subset of these nodes is uniquely determined and can be defined in \FO by the node $w$ and a set $I \subseteq \{1, \ldots, k\}$ that \emph{indexes} the neighbours.
With this idea, the proof of Proposition \ref{prop:dynprop:upperbound} can be adjusted appropriately for Theorem \ref{theorem:parityexists:dynfodegk}.  

\begin{proofof}{Theorem \ref{theorem:parityexists:dynfodegk}}
Let $k \in \N$ be some constant. Again, we call a node \emph{active} if its in-degree is at most $k$. 
We sketch a dynamic program that uses a linear order on the nodes and otherwise at most unary auxiliary relations.

Let $I$ be a non-empty subset of $\{1, \ldots, k\}$, and let $w$ be an active node with at least $\max(I)$ in-neighbours. The set $\Nin_I(w)$ of \emph{$I$-indexed in-neighbours} of $w$ includes a node $v$ if and only if $(v,w)$ is an edge in the input graph and $v$ is the $i$-th in-neighbour of $w$ with respect to the linear order, for some $i \in I$. The following notation is similar as in the proof of Proposition \ref{prop:dynprop:upperbound}. For a graph $G$ and an arbitrary set $C$ of (coloured and uncoloured) nodes, we denote by $\Ncorunc_G(C)$ the set of active nodes that have an incoming edge from every node in $C$ and no coloured in-neighbour that is not in $C$. 
An example for these notions is depicted in Figure~\ref{figure:dynfo:degk}.

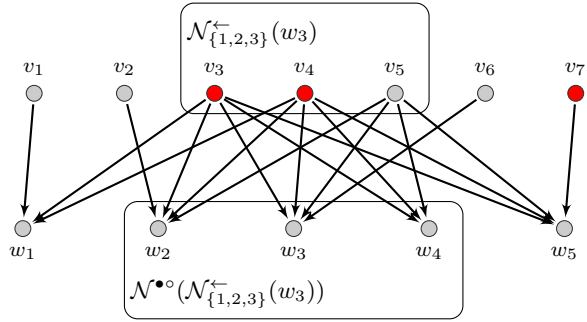
\begin{figure}[t]
\centering
\begin{tikzpicture}[xscale=0.6, yscale=0.6]
\tikzset{rlabel/.style = {font=\scriptsize}}
\tikzset{llabel/.style = {font=\scriptsize}}

\draw [mline, thin] (3.5,-0.45) rectangle (9.0,2);
\draw [mline, thin] (2.25,-2.4) rectangle (9.75,-5);

\node at (5.1,1.3) {\footnotesize $\Nin_{\{1,2,3\}}(w_3)$};
\node at (4.55,-4.45) {\footnotesize $\Ncorunc(\Nin_{\{1,2,3\}}(w_3))$};

\pgfmathsetmacro{\rx}{-3} 
\pgfmathsetmacro{\ry}{0} 
\pgfmathsetmacro{\dist}{3}
\pgfmathsetmacro{\lx}{0} 
\pgfmathsetmacro{\ly}{0.25} 
\pgfmathsetmacro{\ldist}{2}
\node[mnode, label={[llabel]above:$v_1$}] at (\ly+0*\ldist,\lx) (1) {};
\node[mnode, label={[llabel]above:$v_2$}] at (\ly+1*\ldist,\lx) (2) {};
\node[mnode, fill=red, label={[llabel]above:$v_3$}] at (\ly+2*\ldist,\lx) (3) {};
\node[mnode, fill=red, label={[llabel]above:$v_4$}] at (\ly+3*\ldist,\lx) (4) {};
\node[mnode, label={[llabel]above:$v_5$}] at (\ly+4*\ldist,\lx) (5) {};
\node[mnode, label={[llabel]above:$v_6$}] at (\ly+5*\ldist,\lx) (6) {};
\node[mnode, fill=red, label={[llabel]above:$v_7$}] at (\ly+6*\ldist,\lx) (7) {};
\node[mnode, label={[rlabel]below:$w_1$}] at (\ry+0*\dist, \rx) (m1) {};
\node[mnode, label={[rlabel]below:$w_2$}] at (\ry+1*\dist, \rx) (m2) {};
\node[mnode, label={[rlabel]below:$w_3$}] at (\ry+2*\dist, \rx) (m3) {};
\node[mnode, label={[rlabel]below:$w_4$}] at (\ry+3*\dist, \rx) (m4) {};
\node[mnode, label={[rlabel]below:$w_5$}] at (\ry+4*\dist, \rx) (m5) {};

 \path[dEdge] (1) edge (m1);
 \path[dEdge] (3) edge (m1) edge (m2) edge (m3) edge (m4) edge (m5);
 \path[dEdge] (2) edge (m2);
 \path[dEdge] (4) edge (m1) edge (m2) edge (m3) edge (m4) edge (m5);
 \path[dEdge] (6) edge (m3);
 \path[dEdge] (5) edge (m2) edge (m3) edge (m4) edge (m5);
 \path[dEdge] (7) edge (m5);
\end{tikzpicture} 
\caption{An illustration of the notation used in the proof of Theorem~\ref{theorem:parityexists:dynfodegk}. The set $\Ncorunc_G(\Nin_{\{1,2,3\}}(w_3))$ does not include $w_1$, as there is no edge $(v_5,w_1)$, and it does not include $w_5$, as there is an edge $(v_7,w_5)$ for a coloured node $v_7 \not\in \Nin_{\{1,2,3\}}(w_3)$. Compare with Figure~\ref{figure:dynprop:upper}.}\label{figure:dynfo:degk}
\end{figure}

For every $I \subseteq \{1, \ldots, k\}$ with $I \neq \emptyset$ we introduce an auxiliary relation $P_I$ with the following intended meaning. 
An active node $w$ with at least $\max(I)$ neighbours is in $P_I$ if and only if 
\begin{enumerate}[(1)]
\item $w$ has no coloured in-neighbours that are not contained in $\Nin_I(w)$, and 
\item the set $\Ncorunc_G(\Nin_I(w))$ has odd size. 
\end{enumerate}
Note that (1) implies that $w \in \Ncorunc_G(\Nin_I(w))$. %

An auxiliary relation $P_I$ basically replaces the relations $P_{\ell,m}$ with $\ell + m = |I|$ from the proof of Proposition \ref{prop:dynprop:upperbound}, and the updates are mostly analogous.

We now explain how the query relation $\ans$ and the relations $P_I$ are updated when a modification to the input graph occurs. For maintaining $P_I$, observe that first-order formulas can easily express that a node $w$ is active, has at least $\max(I)$ neighbours and satisfies condition (1). The main task of the dynamic program we construct is to check condition (2).

\subsubsection*{Colouring a node $v$}

When a change $\ins_R(v)$ occurs, so, when a node $v$ is coloured, the query relation $\ans$ is only changed if $v$ becomes the only coloured neighbour of an odd number of active nodes. This is the case if and only if there is an active and previously uncovered node $w$ that $v$ has an edge to and, furthermore, if $w \in P_I$ for the set $I \df \{i\}$, where $i$ is the number such that $v$ is the $i$-th in-neighbour of $w$ with respect to the linear order.

For an arbitrary index set $I$, the update of a relation $P_I$ after the colouring of a node $v$ is as follows. Let $G$ be the graph before the change is applied, and let $G'$ be the changed graph.
Let $w$ be any active node. If $v$ is an $I$-indexed in-neighbour of $w$, so if $v \in \Nin_I(w)$, no change regarding $w \in P_I$ is necessary. Otherwise if $v \not\in \Nin_I(w)$, we need to check whether both conditions (1) and (2) from above are satisfied regarding $w$ and $I$. If at least one of the conditions is not satisfied, then $w$ is not contained in the updated relation $P_I$.

Suppose that condition (1) is satisfied, so, that $w$ has no coloured in-neighbours in $G'$ that are not contained in $\Nin_I(w)$. This can easily be checked by a first-order formula. %
We need to check condition (2), so, whether $|\Ncorunc_{G'}(\Nin_I(w))|$ is odd.

The set $\Ncorunc_{G'}(\Nin_I(w))$ contains exactly those nodes from $\Ncorunc_{G}(\Nin_I(w))$ that do not have the newly coloured node $v$ as in-neighbour, as this node is not contained in $\Nin_I(w)$.
So, $\Ncorunc_{G'}(\Nin_I(w)) = \Ncorunc_{G}(\Nin_I(w)) \setminus \Ncorunc_{G}(\Nin_I(w) \cup \{v\})$.
It follows that we can determine the parity of $|\Ncorunc_{G'}(\Nin_I(w))|$ from the parity of $|\Ncorunc_{G}(\Nin_I(w))|$ and the parity of $|\Ncorunc_{G}(\Nin_I(w) \cup \{v\})|$. We know whether $|\Ncorunc_{G}(\Nin_I(w))|$ is odd, this is the case if and only if $w \in P_I$. It remains to find the parity of $|\Ncorunc_{G}(\Nin_I(w) \cup \{v\})|$.

Suppose that $\Ncorunc_{G}(\Nin_I(w) \cup \{v\})$ is non-empty and contains some node $w'$. The node $w'$ is active and has an edge from $v$ and every node in $\Nin_I(w)$, but no edge from any other coloured node. Such a node $w'$ can easily be identified by a first-order formula. Let $I'$ be chosen such that $\Nin_{I'}(w') = \Nin_I(w) \cup \{v\}$.
The set $\Ncorunc_{G}(\Nin_I(w) \cup \{v\})$ coincides with the set $\Ncorunc_G(\Nin_{I'}(w'))$ and the size of this set is odd exactly if $w' \in P_{I'}$. 

So, in this case, condition (2) is satisfied in $G'$ exactly if $P_I(w) \xor P_{I'}(w')$ holds for the old auxiliary relations.
An illustration of the reasoning is given in Figure \ref{figure:dynfo:degkdetail}.

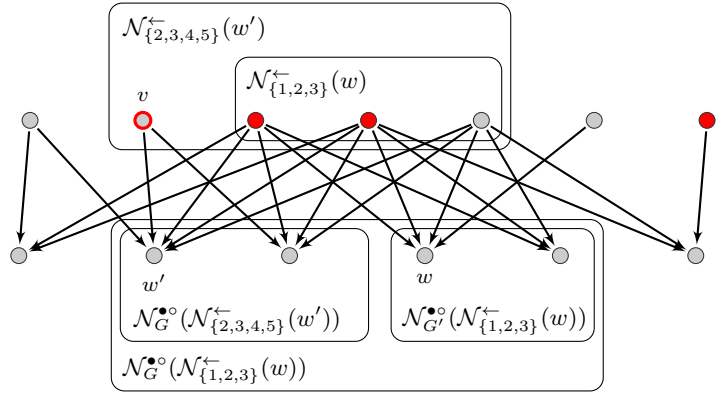
\begin{figure}[t]
\centering
\begin{tikzpicture}[xscale=0.6, yscale=0.6]
\tikzset{rlabel/.style = {font=\scriptsize}}
\tikzset{llabel/.style = {font=\scriptsize}}

\draw [mline, thin] (4.8,-0.45) rectangle (10.7,1.4);
\draw [mline, thin] (2.0,-0.65) rectangle (10.9,2.6);
\draw [mline, thin] (2.05,-2.2) rectangle (12.95,-6);
\draw [mline, thin] (2.25,-2.4) rectangle (7.75,-4.9);
\draw [mline, thin] (8.25,-2.4) rectangle (12.75,-4.9);

\node at (6.4,0.8) {\footnotesize $\Nin_{\{1,2,3\}}(w)$};
\node at (3.9,2) {\footnotesize $\Nin_{\{2,3,4,5\}}(w')$};
\node at (10.55,-4.45) {\footnotesize $\Ncorunc_{G'}(\Nin_{\{1,2,3\}}(w))$};
\node at (4.85,-4.45) {\footnotesize $\Ncorunc_{G}(\Nin_{\{2,3,4,5\}}(w'))$};
\node at (4.35,-5.55) {\footnotesize $\Ncorunc_{G}(\Nin_{\{1,2,3\}}(w))$};

\pgfmathsetmacro{\rx}{-3} 
\pgfmathsetmacro{\ry}{0} 
\pgfmathsetmacro{\dist}{3}
\pgfmathsetmacro{\lx}{0} 
\pgfmathsetmacro{\ly}{0.25} 
\pgfmathsetmacro{\ldist}{2.5}
\node[mnode, label={[llabel]above:$ $}] at (\ly+0*\ldist,\lx) (1) {};
\node[mnode, draw=red, very thick, label={[llabel]above:$v$}] at (\ly+1*\ldist,\lx) (2) {};
\node[mnode, fill=red, label={[llabel]above:$ $}] at (\ly+2*\ldist,\lx) (3) {};
\node[mnode, fill=red, label={[llabel]above:$ $}] at (\ly+3*\ldist,\lx) (4) {};
\node[mnode, label={[llabel]above:$ $}] at (\ly+4*\ldist,\lx) (5) {};
\node[mnode, label={[llabel]above:$ $}] at (\ly+5*\ldist,\lx) (6) {};
\node[mnode, fill=red, label={[llabel]above:$ $}] at (\ly+6*\ldist,\lx) (7) {};
\node[mnode, label={[rlabel]below:$ $}] at (\ry+0*\dist, \rx) (m1) {};
\node[mnode, label={[rlabel]below:$w'$}] at (\ry+1*\dist, \rx) (m2) {};
\node[mnode, label={[rlabel]below:$ $}] at (\ry+2*\dist, \rx) (m3) {};
\node[mnode, label={[rlabel]below:$w$}] at (\ry+3*\dist, \rx) (m4) {};
\node[mnode, label={[rlabel]below:$ $}] at (\ry+4*\dist, \rx) (m5) {};
\node[mnode, label={[rlabel]below:$ $}] at (\ry+5*\dist, \rx) (m6) {};

 \path[dEdge] (1) edge (m1) edge (m2);
 \path[dEdge] (3) edge (m1) edge (m2) edge (m3) edge (m4) edge (m5);
 \path[dEdge] (2) edge (m2) edge (m3);
 \path[dEdge] (4) edge (m1) edge (m2) edge (m3) edge (m4) edge (m5) edge (m6);
 \path[dEdge] (6) edge (m4);
 \path[dEdge] (5) edge (m2) edge (m3) edge (m4) edge (m5) edge (m6);
 \path[dEdge] (7) edge (m6);
\end{tikzpicture} 
\caption{Illustration for the update of the relation $P_I$ in the proof of Theorem~\ref{theorem:parityexists:dynfodegk}, for $I = \{1,2,3\}$. Graph $G'$ results from graph $G$ by colouring node $v$. The set $\Ncorunc_{G}(\Nin_{\{1,2,3\}}(w))$ is the disjoint union of the sets $\Ncorunc_{G}(\Nin_{\{2,3,4,5\}}(w'))$ and $\Ncorunc_{G'}(\Nin_{\{1,2,3\}}(w))$, so the parity of $|\Ncorunc_{G'}(\Nin_{\{1,2,3\}}(w))|$ can be determined from the parity of $|\Ncorunc_{G}(\Nin_{\{2,3,4,5\}}(w'))|$ and the parity of $|\Ncorunc_{G}(\Nin_{\{2,3,4,5\}}(w'))|$.}\label{figure:dynfo:degkdetail}
\end{figure}

Note that if no such $w'$ exists, then $\Ncorunc_G(\Nin_I(w)) = \Ncorunc_{G'}(\Nin_I(w))$ and no change regarding $w \in P_I$ is necessary.

\subsubsection*{Uncolouring a node $v$}

The update of the query relation $\ans$ and the relations $P_I$ after a change $\del_R(v)$ that uncolours a node $v$ are symmetric to the previous case.
The query bit is flipped if there is an active node $w$ that has $v$ as its only coloured in-neighbour and if $w \in P_I$ for the index set $I = \{i\}$, where $i$ is the number such that $v$ is the $i$-th in-neighbour of $w$.
The update of auxiliary relations $P_I$ for some node $w$ is exactly as for the previous case of changes $\ins_R(v)$ --- for checking condition (2) in the case $v \not\in \Nin_I(w)$, we now have that $\Ncorunc_{G'}(\Nin_I(w))$ is the disjoint union of $\Ncorunc_G(\Nin_I(w))$ and $\Ncorunc_G(\Nin_{I'}(w'))$, where $I'$ and $w'$ are defined exactly as in the previous case, still condition (2) holds exactly if $P_I(w) \xor P_{I'}(w')$ holds.

\subsubsection*{Inserting an edge $(v,w)$}

When a change $\ins_E(v,w)$ inserts an edge $(v,w)$, the query bit $\ans$ needs to be flipped if $w$ is a covered node that becomes inactive, so if its degree becomes larger than $k$, or if $w$ stays active and $v$ becomes its first coloured in-neighbour. Otherwise, $\ans$ is not changed.

For any index set $I$, the relation $P_I$ is updated as follows. 
Let $G$ be the graph before the change, and let $G'$ be the changed graph that includes the edge $(v,w)$.
First, we consider whether some active node $u \neq w$ is in the updated version of $P_I$. We assume that condition~(1) from above holds for $u$, otherwise $u \not\in P_I$ before and after the update. 
Whether $u \in P_I$ holds might only change if the parity of $\Ncorunc(\Nin_I(u))$ changes. This only happens when the membership of $w$ in $\Ncorunc(\Nin_I(u))$ changes by inserting $(v,w)$, and this happens if
\begin{enumerate}[(a)]
 \item $w \in \Ncorunc_G(\Nin_I(u))$ and
  \begin{enumerate}[(i)]
   \item $w$ becomes inactive, or
   \item $v$ is a coloured node that is not included in $\Nin_I(u)$; or
  \end{enumerate}
\item $w \notin \Ncorunc_G(\Nin_I(u))$ and $w$ stays active, the node $v$ is in $\Nin_I(u)$, and $w$ has an incoming edge from all nodes from $\Nin_I(u)$ in $G'$ and no incoming edge from any coloured node that is not in this set.
\end{enumerate}

It remains to explain how to determine whether $w \in P_I$ needs to hold after the update. In the following, we make the graph explicit in the notation $\Nin_I(w)$ and write $\Nin_{G,I}(w)$ for the set if $I$-indexed in-neighbours of $w$ in $G$.
We assume that $w$ stays active and that condition (1) from above holds after the change, so $w$ has no coloured in-neighbours that are not in $\Nin_{G',I}(w)$. The update formulas only need to check whether condition (2) holds, that is, whether the set $\Ncorunc_{G'}(\Nin_{G',I}(w))$ has odd size. 

We consider two cases. 
First, suppose $v \not \in \Nin_{G',I}(w)$. As $w$ satisfies condition (1), it follows that $v$ is uncoloured. Let $I'$ be the index set such that $\Nin_{G',I}(w)=\Nin_{G,I'}(w)$. Then $\Ncorunc_{G'}(\Nin_{G',I}(w)) = \Ncorunc_{G}(\Nin_{G,I'}(w))$, and $w \in P_I$ holds after the update if and only if $w \in P_{I'}$ before the update.

Now, suppose $v \in \Nin_{G',I}(w)$. If $w$ is not the only node in $\Ncorunc_{G'}(\Nin_{G',I}(w))$, then there is another active node $w'$ and an index set $I'$ such that $\Nin_{G',I}(w) = \Nin_{G,I'}(w')$ and $w'$ has no coloured in-neighbour apart from nodes in  $\Nin_{G,I'}(w')$. It follows that $\Ncorunc_{G'}(\Nin_{G',I}(w)) = \{w\} \cup \Ncorunc_{G}(\Nin_{G,I'}(w'))$. So, $w \in P_I$ holds after the update if and only if $w' \not\in P_{I'}$ before the update.

All conditions can easily be expressed by first-order formulas.

\subsubsection*{Deleting an edge $(v,w)$}

The update after a change $\del_E(v,w)$ is mostly symmetric to the update after the insertion of an edge $(u,v)$. We shortly describe it here.
The query bit $\ans$ needs to be flipped if $w$ is a covered and formerly inactive node that becomes active, so, if $w$ has degree $k$ after applying the change and has a coloured in-neighbour different from $v$.
The bit $\ans$ also needs to be flipped if $w$ is a covered and active node before applying the change, but $v$ is the only node that covers $w$.
In all other cases, $\ans$ is not changed.

We now discuss the update of $P_I$, for some index set $I$.
Let $G$ be the graph before the change, and let $G'$ be the changed graph that does not include the edge $(v,w)$.
For a node $u \neq w$, the reasoning for updating whether $u \in P_I$ holds is very similar to the reasoning in the case of inserting an edge $(u,v)$. The only differences are the conditions that tell whether the membership of $w$ in $\Ncorunc(\Nin_I(u))$ changes. This happens after the edge $(v,w)$ is deleted, if
\begin{enumerate}[(a)]
 \item $w \in \Ncorunc_G(\Nin_I(u))$ and the node $v$ is in $\Nin_I(u)$, or
\item $w \notin \Ncorunc_G(\Nin_I(u))$, the node $w$ is active in $G'$, has an incoming edge from all nodes from $\Nin_I(u)$ in $G'$, but not from any coloured node that is not in this set, and
   \begin{enumerate}[(i)]
    \item $w$ was not active in $G$, or
    \item $v$ is a coloured node that is not included in $\Nin_I(u)$.
    \end{enumerate}
\end{enumerate}

Now we explain the update of $P_I$ for the node $w$. We assume that $w$ is active in $G'$ and satisfies condition (1) from above; otherwise, $w \not\in P_I$.
We need to explain how to determine the parity of $|\Ncorunc_{G'}(\Nin_{G',I}(w))|$.
We again consider two cases. First, suppose that $w$ was already active in $G$ and that $v$ is uncoloured. 
Let $I'$ be the index set such that $\Nin_{G',I}(w)=\Nin_{G,I'}(w)$. Again, $\Ncorunc_{G'}(\Nin_{G',I}(w)) = \Ncorunc_{G}(\Nin_{G,I'}(w))$, and $w \in P_I$ holds after the update if and only if $w \in P_{I'}$ before the update.

Suppose the first case does not apply. Then, if $w$ is not the only node in $\Ncorunc_{G'}(\Nin_{G',I}(w))$, there is again an active node $w' \neq w$ and an index set $I'$ such that $\Nin_{G',I}(w) = \Nin_{G,I'}(w')$ and $w'$ has no coloured in-neighbour apart from nodes in  $\Nin_{G,I'}(w')$. It holds that $\Ncorunc_{G'}(\Nin_{G',I}(w)) = \{w\} \cup \Ncorunc_{G}(\Nin_{G,I'}(w'))$ and so, $w \in P_I$ after the update if and only if $w' \not\in P_{I'}$ before the update.

Again, all conditions are easily seen to be first-order expressible. 
\end{proofof}

It is easy to maintain a linear order on the non-isolated nodes of an input graph \cite{Etessami98}, which is all that is needed for the proof of Theorem \ref{theorem:parityexists:dynfodegk}. So, $\parityexistsdeg{k}$ can also be maintained in \DynFO without a predefined linear order, at the expense of binary auxiliary relations. 

Unfortunately we cannot generalise the technique from the proof of Theorem~\ref{theorem:parityexists:dynfodegk} for \parityexistsdeg{k} to \parityexists, but only to $\parityexistsdeg{\log n}$, which asks for the parity of the number of covered nodes with in-degree at most $\log n$. Here, $n$ is the number of nodes of the graph.

\begin{customthm}{\ref{theorem:parityexists:dynfologn}}
\statementDynFOlogn
\end{customthm}
\begin{proofsketch}
With the help of the linear order we identify the node set $V$ of size $n$ of the input graph with the numbers $\{0, \ldots, n-1\}$, and use $\BIT$ to access the bit encoding of these numbers.
Any node $v \in V$ then naturally encodes a set $I(v) \subseteq \{1, \ldots, \log n\}$: $i \in \{1, \ldots, \log n\}$ is contained in $I(v)$ if and only if the $i$-th bit in the bit encoding of $v$ is $1$.

The proof of Theorem~\ref{theorem:parityexists:dynfodegk} constructs a dynamic program that maintains unary relations~$P_I$, for each non-empty set $I \subseteq \{1, \ldots, k\}$. We replace these relations by a single binary relation $P$, with the intended meaning that $(v,w) \in P$ if and only if $w \in P_{I(v)}$.
First-order update rules can easily translate between these two representations in the presence of a linear order and $\BIT$, and otherwise the update works exactly as described in the proof of Theorem~\ref{theorem:parityexists:dynfodegk}.
\end{proofsketch}

In addition to a linear order, Etessami \cite{Etessami98} also shows how corresponding relations addition and multiplication can be maintained for the active domain of a structure. As $\BIT$ is first-order definable in the presence of addition and multiplication, and vice versa (see e.g.\ \cite[Theorem~1.17]{ImmermanDC}), both a linear order and $\BIT$ on the active domain can be maintained, still using only binary auxiliary relations. 
So, the variant of $\parityexistsdeg{\log n}$ that considers $n$ to be the number of non-isolated nodes, instead of the number of all nodes, can be maintained in binary \DynFO without assuming built-in relations.

\section{Maintenance using auxiliary relations of quasi-polynomial size}
\label{section:quasi}
Orthogonally to the perspectives taken in this work so far, which focussed on the expressive power of the update formalism, one can ask how many auxiliary bits are necessary to maintain the query \parityexists, or, more generally, all queries expressible in first-order logic extended by modulo quantifiers. 
The class $\DynFO$ allows for polynomially many auxiliary bits: the auxiliary relations of a \DynFO program can be encoded by a bit string of polynomial size. It is not hard to see that if one allows quasi-polynomially many auxiliary bits -- so, the number of auxiliary bits is bounded by $2^{\log^{\bigO(1)} n}$ -- then all queries expressible in first-order logic extended by modulo quantifiers can be maintained. This was observed in discussions with Samir Datta, Raghav Kulkarni and Anish Mukherjee. Here, we provide a proof sketch for this observation.

For discussing the amount of auxiliary bits, it is convenient to switch the view point from first-order updates to updates computed by $\AC^0$ circuits.
A classical result linking circuit complexity and finite model theory states that a query can be computed by a uniform family of $\AC^0$-circuits (that is, by constant depth and polynomial size circuits with $\neg$-, $\wedge$- and $\vee$-gates with unbounded fan-in) if and only if it can be expressed by a first-order formula with access to a built-in linear order and $\BIT$ \cite{BarringtonIS90}. 
So, if we assume the presence of a built-in linear order and $\BIT$ then the classes $\DynFO$ and (uniform) $\DynAC^0$ coincide.

The class $\ACC^0$ is defined similarly as $\ACz$, but the circuits are additionally allowed to use modulo-gates. A query can be computed by a uniform family of $\ACC^0$-circuits if and only if it can be expressed by a first-order formula that may use modulo quantifiers, in addition to a linear order and $\BIT$. 

For simplifying the discussion, in the following we take a solely circuit-based perspective. We also, from now on, disregard uniformity conditions and only consider non-uniform circuit classes.

The classes $\qAC^0$ and $\qACC^0$ are defined as the classes $\AC^0$ and $\ACC^0$ except that circuits can be of quasi-polynomial size, that is, of size $2^{\log^{\bigO(1)} n}$. The class $\DynAC^0$ is the class of queries that can be maintained with $\AC^0$-circuits and polynomially many auxiliary bits. The class $\qDynAC^0$ is defined as the class $\DynAC^0$ except that dynamic programs may use quasi-polynomially many auxiliary bits and update circuits from $\qAC^0$.

It turns out that with quasi-polynomial update circuits all $\qACC^0$-queries can be maintained, and in particular the query \parityexists.

\begin{thm}\label{theorem:QACCzInQDynACZ}
  Every query in $\qACC^0$ can be maintained in $\qDynAC^0$.
\end{thm}

Instead of proving this theorem directly, we use that $\qACC^0$ can be characterised by very simple circuits with one gate with quasi-polynomial in-degree \cite{BeigelT94}.

A boolean function $f: \{0,1\}^m \rightarrow \{0,1\}$ is \emph{symmetric} if $f(\tpl x) = f(\tpl y)$ whenever the number of ones in $\tpl x$ and $\tpl y$ is equal. The class $\Sym$ contains all queries computable by depth-two size-$2^{\log^{\bigO(1)} n}$ circuits where the output gate computes a symmetric boolean function, and it has incoming wires from a layer of and-gates, which each have fan-in $\log^{\bigO(1)} n$ (see Reference \cite{BeigelT94}). 

As $\qACC^0$ is contained in $\Sym$ \cite[Proposition 1.2]{BeigelT94}, the following result implies Theorem \ref{theorem:QACCzInQDynACZ}. 
\begin{thm}\label{theorem:SymInQDynACZ}
  Every query in $\Sym$ can be maintained in $\qDynAC^0$.
\end{thm}
\begin{proof}
The proof extends the ideas from the proofs of Theorem \ref{theorem:parityexists:dynfologn} and Theorem \ref{theorem:parityexists:dynfodegk} to nodes with in-degree $\log^{\bigO(1)} n$.

  Let $\calC$ be a family of $\Sym$-circuits, where for each $n \in \N$ the circuit $C_n$ is a depth-two size-$2^{\log^{\bigO(1)} n}$ circuit whose output gate computes some symmetric boolean function $h$, and that otherwise consists of a layer of and-gates $g_1, g_2, \ldots$ with fan-in at most $k$, for some $k \in \log^{\bigO(1)} n$.
  
  We construct a $\qDynAC^0$-program $\prog$. The idea is to maintain, for each domain size $n$, the number $m$ of and-gates of $C_n$ that are currently \emph{activated}, i.e., that are only connected to inputs that are currently set to 1. The output of the symmetric function $h$ for inputs with $m$ ones can then be looked up in a table.
  
For maintaining the number of activated and-gates, the program $\prog$ maintains, for every subset $A$ of the input gates of size at most $k$, the value 
\begin{align*}
 \#(A) \df |\{g_i \mid \, &\text{$g_i$ is connected to all inputs in $A$,} \\ 
 & \text{and is activated when the inputs from $A$ are ignored}\}|,
\end{align*}
that is, the number of and-gates that are connected to all inputs in $A$ and whose only inputs that are not set to $1$ (if any) are contained in $A$. 

Rephrased in the setting of the previous sections, we can think of the circuit as a coloured graph $G$, and an input gate is considered to be coloured if the input bit is set to $0$. Then, $\#(A)$ is the number of and-gates that are connected to all inputs in $A$ and are \emph{not} covered by some node that is \emph{not} in $A$, so, the cardinality of the set $\Ncorunc_G(A)$ as defined in Section~\ref{section:first-order}.

As the number of sets of input gates of size at most $k$ is quasi-polynomial in $n$, and for each such set $A$ the number $\#(A)$ is bounded by the size of the circuit, which is again quasi-polynomial in $n$, this auxiliary information can be encoded by quasi-polynomially many auxiliary bits.

We need to show that the number of activated gates as well as the auxiliary information $\#(A)$ can be updated after changes that flip an input bit. Note that the circuit itself is fixed, so the dynamic program does not need to support changes that delete or insert wires.
  
  When an input bit $x$ is changed, the values $\#(A)$ are updated to $\#'(A)$ as follows. 
For all sets $A$ that include $x$, the count $\#(A)$ does not change, as all inputs in $A$ are ignored in the definition of $\#(A)$. Also, the count $\#(A)$ does not change for sets $A$ with $x \not\in A$ and $|A| = k$, as all and-gates have fan-it at most $k$ and can therefore not be connected to all inputs in $A$ and the additional input $x$.
The only case left is for a set $A$ with $|A|<k$ that does not include $x$. Suppose that $x$ is flipped from $0$ to $1$.
All and-gates counted for $\#(A)$ are not connected to $x$ by the definition of $\#(A)$. After the change, also those gates are counted for $\#(A)$ that have a connection to all gates in $A$ and to $x$, but to no other input gate that is set to $0$. The number of these gates is given by $\#(A \cup \{x\})$.

In summary, the updated auxiliary information $\#'(A)$ after $x$ is set from $0$ to $1$ is as follows:
    \[ \#'(A) \df   \begin{cases}
                      \#(A)  & \text{if $x \in A$ or $|A| = k$}\\
                      \#(A) + \#(A \cup \{x\}) & \text{else.}\\
                    \end{cases}\]
If $x$ is set from $1$ to $0$, the update is symmetric:
    \[ \#'(A) \df   \begin{cases}
                      \#(A)  & \text{if $x \in A$ or $|A| = k$}\\
                      \#(A) - \#(A \cup \{x\})  & \text{else.}\\
                    \end{cases}\]
The total number of activated gates changes by the number $\#(\{x\})$, which is either added in case $x$ is flipped from $0$ to $1$, or subtracted if $x$ is flipped from $1$ to $0$. These updates can easily be expressed by $\qAC^0$ circuits.             
\end{proof}

In Reference \cite{Mukherjee19} it is discussed how Theorem \ref{theorem:QACCzInQDynACZ} can be extended to show that all queries in $\qAC$ can be maintained in $\qDynAC^0$, using different techniques. Here, $\qAC$ denotes the class of all queries that can computed by families of circuits with quasi-polynomial size and poly-logarithmic depth. 
\section{Conclusion}
\label{section:conclusion}

We studied the dynamic complexity of the query \parityexists as well as its bounded degree variants. While it remains open whether \parityexists is in $\DynFO$, we showed that $\parityexistsdeg{\log n}$ is in $\DynFO$ and that $\parityexistsdeg{k}$ is in \DynProp, for fixed $k \in \N$. The latter result is the basis for an arity hierarchy for $\DynProp$ for Boolean graph queries. Several open questions remain.

\begin{openquestion*}
 Can {\parityexists} be maintained with first-order updates rules? If so, are all (domain-independent) queries from \FOparity also in $\DynFO$?
\end{openquestion*}

\begin{openquestion*}
 Is there an arity hierarchy for $\DynFO$ for Boolean graph queries?
\end{openquestion*} 
\section*{Acknowledgment}
  \noindent This project has received funding
from the European Union's Horizon 2020 research and innovation
programme under grant agreement No 682588.

\bibliographystyle{alpha}  
\bibliography{bibliography}

\end{document}